\documentclass[final,aps,showpacs,fleqn,nofootinbib,twocolumn]{revtex4-1}
\usepackage[toc]{appendix}
\usepackage{epsfig,graphicx}
\graphicspath{./}
\usepackage{amsmath}
\usepackage{hyperref}
\hypersetup{colorlinks,linkcolor=red, citecolor=green}

\newcommand{\be}{\begin{equation}}
\newcommand{\ee}{\end{equation}}
\newcommand{\bea}{\begin{eqnarray}}
\newcommand{\eea}{\end{eqnarray}}
\newcommand{\ba}{\begin{array}}
\newcommand{\ea}{\end{array}}
\newcommand{\tr}{{\rm tr}}

\usepackage[normalem]{ulem} % \sout{old text} for strikeout
\usepackage[dvipsnames]{xcolor} % For blue in-text comments and
\renewcommand\sout{\bgroup \color{red} \ULdepth=-.5ex \ULset}

\usepackage{tikz-feynman}
\usetikzlibrary{shapes,arrows,positioning,automata,backgrounds,calc,er,patterns}
\tikzfeynmanset{compat=1.0.0, every vertex={dot,minimum size=2pt}}
\begin{document}

\title{Heavy quark correlators in 
Instanton Liquid Model with perturbative corrections}
\author{M.\,Musakhanov}
%\email{yousufmm@list.ru}
\affiliation{
National University of Uzbekistan, Tashkent 100174, 
Uzbekistan} 
\author{N.\,Rakhimov}
\affiliation{
National University of Uzbekistan, Tashkent 100174, 
Uzbekistan} 
\author{U.T.\,Yakhshiev}
\affiliation{
Inha University, Incheon 22212, Republic of Korea} 
\affiliation{
National University of Uzbekistan, Tashkent 100174, 
Uzbekistan} 
\affiliation{
Institute of Nuclear Physics, AS RUz, Tashkent 100214, 
Uzbekistan} 

\date\today

\begin{abstract}

In the present work we consider the influence on the heavy quark correlators
due to the instanton background  in the framework of 
instanton liquid model (ILM) of  
QCD vacuum by taking into account also 
the perturbative gluon effects.
For a single heavy quark this leads to the mass shift due to 
the direct-instanton nonperturbative and ILM modified 
perturbative contributions, respectively.
In the  heavy quark-antiquark 
($Q\bar{Q}$)  sector we obtain the potential
consisting the direct instanton induced part and the 
one-gluon exchange (OGE) perturbative part which is screened 
at large distances due to the nonperturbative dynamics.
At the region of interest corresponding to the heavy quark physics
the screening effect in OGE can be well approximated by a Yukawa-type
potential in terms of the dynamically generated gluon mass.
 A possible implication of the present studies to
the phenomenology of heavy quarkonium is also discussed.

\end{abstract}


\maketitle
\thispagestyle{empty}

\section{Introduction}

The properties of heavy quarkonium $Q\bar Q$ 
(a colorless system consisting a heavy quark $Q$ and another heavy antiquark $\bar Q$) 
in the framework 
of phenomenological approaches can be well described 
essentially on a basis of non-relativistic potential models. 
The  model quite popular among them is one with so-called
Cornell potential~\cite{Eichten:1974af}, which has a nature of 
Coulomb-like attractive part at short 
distances and linear confining part at long distances.
The form of potential is given as
\begin{align}
V_{\rm Cornell}(r)=\frac{\kappa}{r}+\sigma r\,,
\label{cornell}
\end{align}
where the Coulomb coupling $\kappa<0$ and 
the string constant $\sigma>0$.
The short-distance behavior of quantum chromodynamics (QCD) is dominated 
by the one-gluon exchange interaction and the corresponding potential can be calculated 
perturbatively. At the leading order on the strong 
coupling constant $\alpha_s=g^2/(4\pi)$, one reproduces 
the first term in Eq.\,(\ref{cornell}) with
the constant $\kappa =-(4/3)\alpha_s$
where `$-4/3$' is color factor corresponding to
 the color singlet $Q\bar{Q}$ state. 
Nevertheless, in the practical calculations $\kappa$ is mostly considered as a pure phenomenological parameter.

The confining part of the potential is
 pure phenomenological one due to the fact that 
we do not fully understand yet the mechanism of confinement. Therefore, some authors 
are using the harmonic oscillator type $\sim r^2$ or the 
logarithmic $\sim \ln (r)$ dependencies for the confining part of potential.
From other side, the lattice QCD calculations showed the linear $\sim r$ dependence of the full potential
at large distances~(see, e.g. Ref.\,\cite{Bali:2000gf})
supporting in such a way the Cornell-type form of
potentials. Actually, all these potential models 
with the different confining forms
reasonable well match 
with the data due to the 
reason that they do not much affect at 
short distances where we have the sensitive probes.  
This kind of common 
behavior of different confining potentials at the short 
distances is also partial reason for considering 
the Coulomb-coupling  $\kappa$ as a pure 
phenomenological parameter.
 
In principle, one can further try to develop the potential 
approach and improve the description
of data by taking into account the 
relativistic and perturbative corrections on the strong 
coupling constant $\alpha_s$ of QCD~\cite{Brambilla:2009bi,Mateu:2018zym}. 
However, the more influential effects may still be hidden 
behind of the non-perturbative 
structure of QCD vacuum. 
Therefore, one should develop the 
more general systematic approach which takes into account 
not only the next order perturbative effects but also  the 
nonperturbative effects in the properties of heavy quark 
systems. 

Here it is necessary to note, that 
the light quark sector of hadronic 
physics is successfully described by means of spontaneous 
breakdown of chiral symmetry in the framework of instanton 
liquid model of QCD vacuum (ILM)~(see reviews~\cite{Diakonov:2002fq,Schafer:1996wv,shuryak2018}
 and Refs.~\cite{Goeke:2007bj,Goeke:2007nc,Goeke:2010hm,Musakhanov:2012zm,Musakhanov:2018sdu}).  One may also expect 
that the non-perturbative effects on the heavy quark 
propagation in the instanton medium
still may be substantial at last but not least. Consequently, 
our aim in the present work is development of the 
systematic accounting scheme for the 
instanton effects during the calculations of heavy quark 
correlators.  By doing this, we will work in the framework of the 
ILM and simultaneously consider
the perturbative gluon contributions. 

The paper is organized in the following way. In the 
next Section~\ref{sec:ILM} we discuss an applicability
of ILM in the heavy quark sector 
by comparing the parameters of ILM with the 
quark core sizes of hadrons.  Then in 
Section~\ref{sec:QuarkCor}, we explain the formalism 
of calculations of the heavy quark correlators in 
the instanton medium by accounting also the perturbative
gluon corrections.  Section~\ref{subsec:HQP} is devoted
to the calculations of heavy quark propagator in ILM and
estimations of the possible contributions to the heavy quark
mass. Further, two body heavy quark-antiquark correlator 
in ILM 
is presented in Section~\ref{sec:2Qcor}.
In particular, the results 
corresponding to the heavy quark potential are discussed
in Subsections~\ref{subsec:dir} and \ref{subsec:1gluon}.
In Section~\ref{sec:InsEffects} we will analyse the order 
of instanton effects in the quarkonium spectra and will make the
corresponding conclusions.
Finally, in Section~\ref{sec:SumOut}
the present studies and the outlook for future
investigations will be summarized. 
The details of the calculations   of the heavy quark correlators  with the perturbative
gluon corrections in 
ILM
are given in the Appendix~\ref{appendix}.

\section{ILM parameters vs hadron quark core sizes}
\label{sec:ILM}

QCD vacuum has the rich topological structures and, 
probably,  
the most important among them  is an instanton
 --  a classical solution of Yang-Mills equations in 
the 4-dimensional Euclidean space. 
The vacuum of QCD has a degenerate and 
periodic structure 
in the functional space along the collective coordinate 
direction which is 
called the Chern-Simons (CS) coordinate.
Therefore, QCD vacuum can be considered as the
lowest  energy quantum state of the one-dimensional crystal 
along the CS coordinate~\cite{FJR1976,Jackiw:1976pf}.
The QCD instanton (anti-instanton) is a tunneling 
mechanism
in forward (backward) direction between the different 
Chern-Simons states
corresponding to the degenerate vacuum~\cite{Belavin:1975fg}.
The (anti)instanton is described by its collective
coordinates denoted as $\xi_{I}$: the position in 
4-dimensional Euclidean 
space $z_{I}$, the instanton size $\rho_{I}$ and the 
SU$(N_{c})$ color orientation 
given by the unitary matrix $U_{I}$,
$4N_{c}$ variables altogether.\footnote{Hereafter, we drop 
the subscript $I$ for the convenience
and note that $N_c$ is the number of 
colors.} 
The main parameters of ILM 
are the average instanton size
$\bar\rho$ and the inter-instanton distance $R$, or 
the density of instanton media $N/V\equiv
1/R^4$ given in terms of the inter-instanton 
distance\footnote{Here $N$ is the 
total number of instantons.}
(see reviews~\cite{Diakonov:2002fq,Schafer:1996wv,shuryak2018}).
Phenomenologically, their values were estimated as 
\begin{equation}
\bar\rho=\frac{1}{3}\,{\rm fm},\qquad R=1\,{\rm fm}.
\label{eq:SetInstanton}
\end{equation}
These values were found to be in general reasonably and 
confirmed by the theoretical variational calculations~\cite{Diakonov:2002fq,Schafer:1996wv,shuryak2018}
and the lattice simulations of the QCD vacuum~\cite{Chu:1994vi,Negele:1998ev,DeGrand:2001tm,Faccioli:2003qz}.
%%%%%%%%%%%%%%%%%%%%%%%%%%%
\begin{figure}[hbt]
%\vskip 0.5cm
\begin{center}
\includegraphics[scale=0.75]{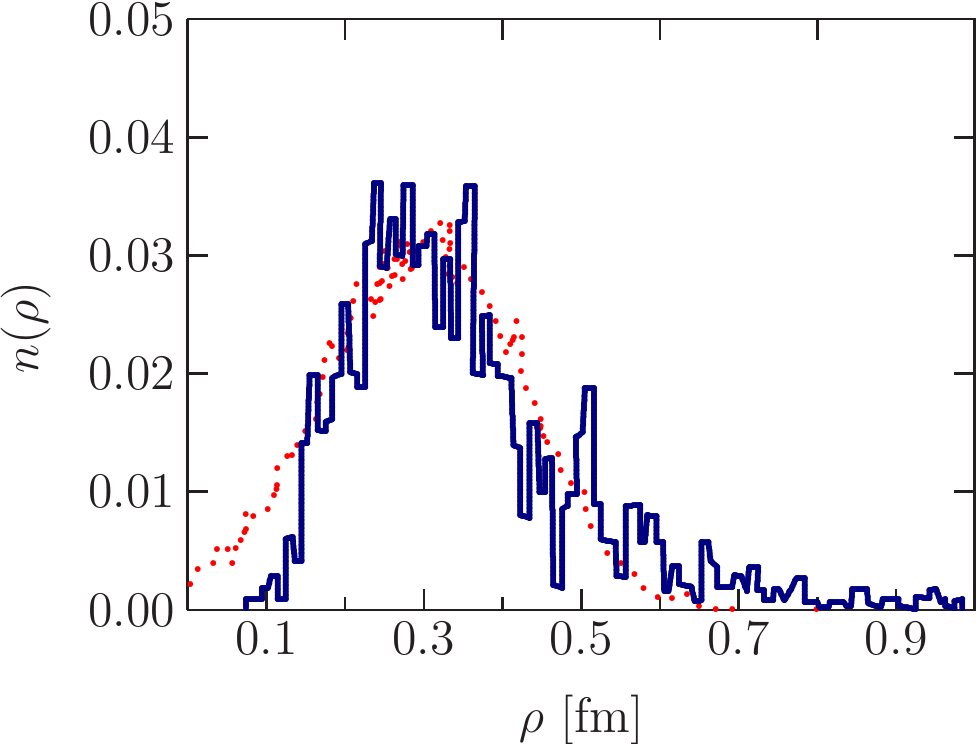}
\end{center} 
\caption{(Color online) Dependence of the instanton size distribution 
function $n$ on the instanton size parameter $\rho$:
the dots correspond to the ILM and the continuous lines
correspond to the lattice simulations~\cite{Millo:2011zn}.}
\label{instantonsize} 
\end{figure}
%%%%%%%%%%%%%%%%%%%%%%%%%%%

From other side, 
the instanton size distribution $n(\rho)$
 has also been studied by lattice simulations~\cite{Millo:2011zn} (see Fig.\ref{instantonsize}).
As we can see, for 
the relatively large-size instantons 
($\rho\sim R$) the distribution function $n(\rho)$ is
suppressed.  Therefore,  a simple sum-ansatz 
for the total instanton field $A(\xi)=\sum_iA_i(\xi_i)$
expressed in terms of the single instanton
solutions  $A_i(\xi_i)$ is quite well justified
during the practical applications. 
Nevertheless, the  the large-size tail of distribution function 
$n(\rho)$ becomes important in the confinement regime of 
QCD. Here one should replace BPST instantons by KvBLL 
instantons~\cite{Kraan:1998kp,Kraan:1998pm,Lee:1998bb}
 described in terms of  dyons. 
In such a way, one gets an extension of ILM \textendash{}
liquid dyon model (LDM)\,\cite{Diakonov:2009jq,Liu:2015ufa,Liu:2015jsa}, which is
able to reproduce confinement\textendash deconfinement
phases. 
Consequently, the small size instantons can still be described in terms of their collective coordinates. For comparison, we remind that
the average size of instantons in LDM is 
$\bar{\rho}\approx0.5\,{\rm fm}$
\cite{Diakonov:2009jq,Liu:2015ufa,Liu:2015jsa}, while in ILM 
$\bar{\rho}\approx0.3\,{\rm fm}$.
Hereafter, we will neglect the effect of 
size distribution's width 
and simply consider the instanton size $\rho$
equal to its average value, $\rho=\bar\rho$.

At the typical values of the ILM parameters
given ~in Eq.\,(\ref{eq:SetInstanton}), the instanton 
background leads to the nonzero
QCD vacuum energy density $\epsilon\approx-500\,
{\rm MeV/fm^3}$\,\cite{Schafer:1996wv,shuryak2018}
and it occurs a spontaneous breakdown of chiral symmetry
which plays the pivotal and significant role in describing the 
lightest hadrons and their interactions. 
In such a way, ILM succeeded to explain 
the hadron physics 
at the light quark sector 
(see  
reviews~\cite{Diakonov:2002fq,Schafer:1996wv,shuryak2018}
and Refs.~\cite{Goeke:2007bj,Goeke:2007nc,Goeke:2010hm,Musakhanov:2012zm,Musakhanov:2018sdu}).

In order to understand the applicability of ILM 
in the heavy quark sector, one
may pay an attention that the typical sizes of quarkonia are 
relatively small~\cite{Digal:2005ht,Eichten:1979ms}.
In particular, the smallness in size is more pronounced in the case of
low laying states, e.g. see  $r_{J/\psi}$ and $r_{\Upsilon}$
in the Table~\ref{Quarkoniumsizes}. 
\begin{table}[h]
\begin{center}
\caption{Masses and  sizes of quarkonium states
in the non-relativistic potential model
\cite{Digal:2005ht}.}
\end{center}
 \scalebox{1}{ %
\begin{tabular}{|c|c|c|c|c|c|c|c|c|}
\hline 
Characteristics
&\multicolumn{3}{c|}{Charmonia}
&\multicolumn{5}{c|}{Bottomonia}\\
\cline{2-9}
of states & $J/\psi$  & $\chi_{c}$  & $\psi'$  & $\Upsilon$  & $\chi_{b}$ & $\Upsilon'$ & $\chi_{b}'$ & $\Upsilon^{''}$ \tabularnewline
\hline 
mass {[}GeV{]}  & 3.07  & 3.53  & 3.68  & 9.46  & 9.99 & 10.02 & 10.26 & 10.36 \tabularnewline
\hline 
size $r$ {[}fm{]}  & 0.25  & 0.36  & 0.45  & 0.14  & 0.22 & 0.28 & 0.34 & 0.39 \tabularnewline
\hline 
\end{tabular}} 
\label{Quarkoniumsizes}
\end{table}
A similar estimate of nucleon's quark core size gives 
the result $r_{N}\sim0.3-0.5$\,fm~\cite{He:1986yq,Weise,Tegen}. 
Due to the fact, that the quark core of hadrons
are relatively small in size, 
one may conclude that they are insensitive to 
the confinement mechanism. 
Consequently, ILM may be safely applied for 
the description of hadron properties at
the heavy quark sector too. However,
one should take care about the perturbative effects
during the analysis of heavy hadrons' spectra
and we will keep it in mind during our calculations. 

\section{Heavy quark correlators with perturbative  corrections}
\label{sec:QuarkCor}

As we discussed above, in ILM the background field due to 
instantons is given by a simple sum 
$A(\xi)=\sum_iA_i(\xi_i)$, where  $\xi_i=(z_i,U_i,\rho_i)$ 
are collective coordinates of instantons. It is also necessary to 
note 
that the instanton field has specific strong coupling dependence given as $A\sim 
1/g$. 
The corresponding partition function in ILM $Z[j]$ 
(it is normalized as $Z[0]=1$) with account of perturbative 
gluons $a_\mu$ and their sources $j_\mu$ is approximated 
by the expression
\begin{align}
Z[j] &= \int  D\xi Da  e^{-[S_{eff}[a,A(\xi)]+(ja)]}\cr
&\approx \int  D\xi e^{-\frac{1}{2}(j_\mu S_{\mu\nu}(\xi)j_\nu)},
\label{Z}
\end{align}
 where it is  neglected the self-interaction terms  at the order 
 of  ${\cal O}(a^3,a^4)$ and used the following
 definitions 
 \begin{align}
 (ja)&=\int d^4x j^a_\mu(x) a^a_\mu(x), \,\,\,\cr
 (j_\mu S_{\mu\nu}(\xi)j_\nu)&=\int d^4xd^4y 
 j^a_\mu(x) S^{ab}_{\mu\nu}(x,y,\xi)j^b_\nu(y).\nonumber
\end{align} 
Here $S^{ab}_{\mu\nu}(x,y,\xi)$ is a gluon propagator in the 
presence of the instanton background $A(\xi)$.
The measure of integration 
in ILM is given as $ D\xi=\prod_i d\xi_i=V^{-1}
\prod_i dz_{i}dU_i$, while the  instantons' sizes $\rho_i$  due to the
inter-instantons interactions are concentrated
around the average size $\bar\rho$. Therefore, in 
ILM for the simplicity it is used 
$\rho_i=\bar\rho$. 

An infinitely heavy quark interacts only 
through the fourth components of 
instantons 
$A_4$ and perturbative gluon $a_4$ fields,
respectively.  In this case we need only $ S_{44}(\xi)$ 
components of a gluon propagator.
 Hereafter, we follow the definitions given in
 Ref.~\cite{Diakonov:1989un}, i.e. $\theta$ is inverse of
 differentiation operator $\theta^{-1}=d/dt$,
  $\langle t|\theta|t'\rangle
 =\theta(t-t')$ is a step-function.  
We also use the following re-definitions of
fields $a\equiv ia_4$, $A\equiv iA_4$, 
source $j\equiv ij_4$ and 
gluon propagator $ S(\xi)\equiv S_{44}(\xi)$. 

With these definitions and re-definitions
the corresponding heavy quark $Q$ and antiquark $\bar Q$ Lagrangians 
can be written as 
\begin{align}
L_Q&=Q^+(\theta^{-1}-ga-gA+...)Q,\\
L_{\bar Q}&=\bar Q^+(\theta^{-1}-g\bar a-g\bar A+...)\bar Q,
\end{align}
where the next order in the inverse of heavy quark mass 
terms are denoted by dots.
In terms of SU($N_c$) generators the quantities
$a$ and $\bar{a}$ are given as 
$a=a_a\lambda_a/2$ and  $\bar a_a=a_a\bar\lambda_a/2$,
where $\bar\lambda_a=-\lambda^{\rm T}_a$.\footnote{Here 
the regular superscript `T' means the operation of transposition.} 
The same rule holds for the fields $A$ and $\bar A$.

During our calculations we will neglect by the virtual processes 
$Q\rightarrow QQ \bar Q$ corresponding to the heavy quark loops
(i.e. the heavy quark determinant equals to 1). 
In such a way the functional space of heavy quarks $Q$ 
is not overlapping with the functional space of heavy 
antiquarks $\bar Q$. Consequently,
the total functional space is a direct product of $Q$ and $\bar 
Q$ spaces.

Let us first consider the simplest heavy quark correlator in 
our approach, i.e. the heavy quark propagator in ILM. 
From Eq.~(\ref{Z}) it is seen, that the averaged heavy quark propagator ${w}$ with the account of perturbative gluon field fluctuations $a$ is given by the expression
\begin{align}
 {w}&=\int D\xi Da \exp[-S_{eff}(a,\xi)+(ja)]\cr
 &\times \left(\theta^{-1}-ga -g\sum_i A_i\right)^{-1}\cr
&=\int D\xi  \left[\int \left(\theta^{-1}-g\frac{\delta}{\delta j} -g\sum_i A_i\right)^{-1}\right.\cr
&\times\left.\exp\left\{\frac{1}{2}(jS(\xi)j)\right\}\right]_{j=0}.
\label{w1}
\end{align}
It is easy to prove that
\begin{align}
&\left[\frac{1}{ \theta^{-1}- g\frac{\delta}{\delta j} -gA(\xi)}
\exp\left(\frac{1}{2}jS(\xi)j\right)\right]_{_{j=0}}\cr
&=
\left[\exp\left(\frac{1}{2}\frac{\delta}{\delta a_a}S_{ab}(\xi)\frac{\delta}{\delta a_b}\right)
\frac{1}{ \theta^{-1}-ga -gA(\xi)}\right]_{_{a=0}}\qquad
\label{q1}
\end{align}
This equation can be extended to any correlator.
Consequently,  the QCD path integral of 
some heavy quark functional $F[A(\xi),a]$  in the approximations discussed above  
can be given by the following equation
\begin{align}
&\int D\xi Da \exp\left\{-S_{eff}[A(\xi),a]\right\} F[A(\xi),a]\cr
&= \int D\xi \left[\exp\left(\frac{1}{2}\frac{\delta}{\delta a_a}S_{ab}(\xi)\frac{\delta}{\delta a_b}\right) F[\xi,a] \right]_{a=0}.
\label{F}
\end{align}
Similar equation in the absence of instanton background 
$A(\xi)=0$ and for the gluon propagator taken in 
Coulomb gauge was suggested before in Ref.~\cite{brown1979}.

The systematic accounting of the nonperturbative effects in the 
ILM can be performed in terms of
the dimensionless parameter, so called the packing parameter 
$ \lambda =\rho^4/R^4$, using the Pobylitsa equations~\cite{Pobylitsa:1989uq}.
The situation here is quite comfortable since in ILM the 
expansion parameter $\lambda$ is very small at the values of instanton 
parameters discussed above, 
$\lambda \sim 0.01$ (see Eq.\,(\ref{eq:SetInstanton})).

It is obvious, that the perturbative effects are taken 
into account in terms of the expansion 
parameter $\alpha_s$. Its
behavior is well established only at the perturbative region
and remains poorly know at the nonperturbative region. 
It is also clear, that the pure perturbative effects at the 
leading order appears as $\sim \alpha_s$,  e.g. the
Coulomb-like interaction part in Eq.\,(\ref{cornell}).  

A systematic analysis including the both perturbative and nonperturbative 
effects requires a double expansion series in terms of $\alpha_s$ and $\lambda$. 
In order to perform such an analysis we assume that 
$\alpha_s\sim \lambda^{1/2}$ which is quite reasonable according to 
the phenomenological studies.
Consequently, we will keep the terms at the order of ${\cal O}(\lambda)$  and
${\cal O}(\alpha_s\lambda^{1/2})$ during our 
calculations. 

At this approximation the gluon propagator in instanton 
medium is represented by re-scattering series as
$$
S(\xi)=S^0+\sum_i \Delta S^i(\xi_i),\quad \Delta S^i(\xi_i)\equiv S^i(\xi_i)-S^0,
$$
where $S^0$ is free and  $S^i(\xi_i)$ is single instanton background gluon propagators,
respectively.
The averaged value of gluon propagator $\overline{S}$ in ILM
can be found by the extension of the Pobylitsa's 
equation to the gluon case~\cite{Musakhanov:2017erp}. It has the form
\begin{align}
\overline{S}(k)= \frac{1}{k^2+M_g^2(k)}, 
\label{glprop}
\end{align}
where the momentum dependent gluon mass is defined
by the following expressions
\begin{align}
M_g(k)&=M_g(0)F(k),\cr
M_g(0)&=\frac{2\pi}{\rho}\left(\frac{6\lambda}{N_c^2-1}\right)
^{1/2}, 
\label{Mg0}\\
\quad F(k)&=k\rho K_1(k\rho).\nonumber
\end{align}
Here $K_1$ is the modified Bessel function of the second type. At the typical 
values of instanton parameters $\rho=1/3\, {\rm fm},\,\,R=1\, {\rm fm}$ one can estimate 
the dynamical gluon mass at zero momentum,
$M_g(0)\simeq 358\,{\rm MeV}$ which is close to the value of the dynamical light quark mass.
One can see, that the dynamical gluon and light quark masses appears 
at the order of ${\cal O}\big(\lambda^{1/2}\rho^{-1}\big)$. The  gauge invariance of $M_g$ was proven in Ref.~\cite{Musakhanov:2017erp}.

 Instantons also generate
the nonperturbative gluon-gluon interactions 
which will contribute to the glueballs' properties.
The investigations in  
ILM~\cite{Schafer:1994fd,Tichy:2007fk} of
the $J^{PC}=0^{++}, 0^{-+},2^{++}$ glueballs, were 
demonstrating that the
instanton-induced forces between gluons lead to the strong
attraction in the $0^{ ++}$ channel, to the 
strong repulsion in the 
$0^{- +}$ channel and to the absence of
short-distance effects in the $2^{ ++}$ channel. 
In such a way, ILM predicted 
hierarchy of the masses 
$m_{0^{++}}<m_{2^{++}}<m_{0^{-+}} $ and their sizes 
$r_{0^{++}}<r_{2^{++}}<r_{0^{-+}} $, 
which were confirmed by
the lattice calculations~\cite{deForcrand:1991kc,Weingarten:1994vc,Chen:1994uw,Morningstar:1999rf,Athenodorou:2020ani,Meyer:2004jc,Meyer:2004gx}.
At typical values of ILM 
parameters $\rho=1/3$~fm and  $R=1$~fm 
there were found~\cite{Schafer:1994fd}, that the mass of $0^{++}$ 
glueball $m_{0^{++}}=1.4\pm 0.2$~GeV and  its size  
$r_{0^{++}}\approx 0.2$~fm in a nice correspondence with
the lattice calculations~\cite{deForcrand:1991kc,Weingarten:1994vc,Chen:1994uw}.
Further studies of the $0^{++}$ glueball in ILM~\cite{Tichy:2007fk} gave 
$m_{0^{++}}=1.29\, -\, 1.42$~GeV, which was also 
in a good agreement with the lattice 
results~\cite{Meyer:2004jc,Meyer:2004gx}.
 
Main conclusion of the works we discussed above
was that the origin of $0^{++}$  glueball is mostly provided by the short-sized 
nonperturbative fluctuations (instantons), rather than the confining forces.

Summarizing all said above, we may conclude that 
ILM provides the consistent framework for describing the
gluon and  the lowest state glueball's properties.

\section{Heavy quark propagator}
\label{subsec:HQP}

An averaged infinitely heavy quark $Q$ propagator in ILM 
according to Eqs.(\ref{w1})-(\ref{q1}) is given as 
\begin{eqnarray}\label{w}
 {w}=\!\left.
\!\int\!\! D\xi \exp\left[\frac{1}{2}\!
\left(\frac{\delta}{\delta a}S(\xi)\frac{\delta}{\delta a}\right)\!\right] \!
\frac{1}{\theta^{-1}-ga -gA(\xi)}  \right|_{a=0}\!\!\!
\end{eqnarray}
where
\begin{equation}
\left(\frac{\delta}{\delta a}S(\xi)\frac{\delta}{\delta a}\right)=
\int d^4yd^4z
\frac{\delta}{\delta a_a(y)}S_{ab}(\xi,y,z)\frac{\delta}{\delta a_b(z)}.
\end{equation}
The details of systematic analysis of the heavy quark propagator is discussed in Appendix~\ref{appendix}.  
From Eq.~(\ref{Qprop}) we see that the heavy quark $Q$ propagator in ILM with perturbative corrections 
can be written as
\begin{equation}
 {w}=\int D\xi\left[\theta^{-1} -\sum_i \left( g A_i(\xi_i) -g^2\left( \Delta S^i(\xi_i)\theta\right)\right)\right]^{-1},
\label{w11}
\end{equation}
where the last term in the denominator of (\ref{w11}) has a meaning of ILM perturbative the heavy quark mass operator of the order ${\cal O}(\alpha_s\lambda^{1/2})$.

Heavy quark propagator Eq.~(\ref{w11}) and its $g\rightarrow 0$ limit expression have the similar structures according to their dependencies on the instanton collective coordinates.
Consequently, we may easily extend Pobylitsa equation in Ref.~\cite{Diakonov:1989un} 
for our purpose. The corresponding extension in 
the approximation ${\cal O}(\lambda,\alpha_s\lambda^{1/2})$ has form
\begin{eqnarray}
 {w}^{-1}&=&\theta^{-1}-\sum_i \int d\xi_i \Big[\theta^{-1} (\frac{1}{ \theta^{-1} - gA_i(\xi_i)}-\theta)\theta^{-1}\cr
&+&g^2\left(\Delta S^i(\xi_i)\theta\right)\Big]
\nonumber\\
&=&\theta^{-1}-\sum_i \int d\xi_i 
\theta^{-1}\left(\frac{1}{ \theta^{-1} -g A_i(\xi_i)}-\theta\right)\theta^{-1} \cr
&-&g^2 \left((\bar S-S^0) \theta\right).
\label{w-1}
\end{eqnarray}
In the last term the averaged 
gluon propagator $\bar S$ is given by Eq.~(\ref{glprop}).
In such a way, the second term 
in the right side of Eq.~(\ref{w-1}) leads to the ILM heavy 
quark mass shift $\Delta M_Q^{\rm dir}$ with the corresponding 
order  ${\cal O}(\lambda)$ while the third one is ILM modified  
perturbative gluon contribution to the heavy quark mass 
$\Delta M_Q^{\rm pert}$ with order of  ${\cal O}(\alpha_s\lambda^{1/2})$, respectively. 

The more detailed discussions of the perturbative contributions to the heavy quark 
mass in ILM $\Delta M_Q^{\rm pert}$ are given in the Appendix~\ref{appendix} (e.g., 
see Eq.~(\ref{DMpert1})) while  the direct instanton one $\Delta M_Q^{\rm dir}$ 
was given in 
Ref.~\cite{Diakonov:1989un}. Finally, we have the following estimation 
 $$
\Delta M_Q^{\rm pert}\le 
\frac{2}{N_c}\,\alpha_s M_g(0)
\sim  \Delta M_Q^{\rm dir}\simeq 70\, {\rm MeV} 
$$
at $N_c=3$
for the given values of model parameters: $\alpha_s= 0.3,\,
\rho=1/3\, {\rm fm}$, $R=1\, {\rm fm}$. This estimation is in accordance with our assumptions ${\cal O}(\alpha_s\lambda^{1/2})\sim {\cal O}(\lambda)$ and
shows that the instanton-perturbative gluon interaction accounts 
the non-negligible changes of the perturbative 
gluon corrections.

\section{$Q\bar Q$ correlator}
\label{sec:2Qcor}

\begin{figure}[hbt]
\vskip 0.5cm
\includegraphics[width=7cm,angle=0]{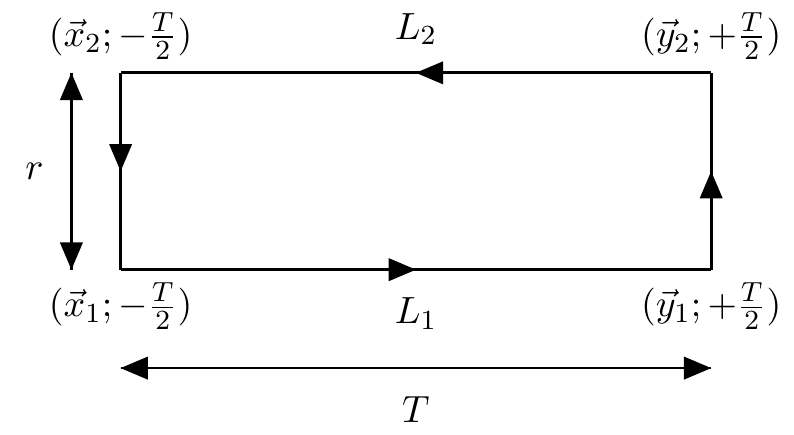}
\caption{The rectangular Wilson loop
with the long time-direction sides $T$ 
and the short space direction sides $r$. }
\label{Wilsonloop}
\end{figure}

In the ILM the heavy quark-antiquark correlator 
 $\langle Q\bar Q\rangle $ is given by the equation
\begin{eqnarray}
&&\langle \bar x, a, x, b|W|\bar x', d, x', g\rangle \cr
&&=\langle 0|T\bar Q_a(\bar x)Q_b(x)Q^+_g(x')\bar Q^+_d(\bar x')|0\rangle\cr
&&=\delta^3(\vec x-\vec x')\delta^3(\vec{\bar x}-\vec{\bar x'})
\theta(t-t')\theta(\bar t-\bar t') \cr
&&\times\int D\xi Da\exp(-S_{eff}(a,\xi))\cr
&&\times\left[T\exp\left(ig\int_{t'}^t d\tau(a(\vec x,\tau)+A(\xi,\vec x,\tau))\right)\right]_{b g}\cr
&&\times\left[T\exp\left(ig\int_{\bar t'}^{\bar t} 
d\bar\tau(\bar a(\vec{\bar x},\bar\tau)+\bar A(\xi,\vec{\bar x},\bar\tau))\right)\right]_{a d},\qquad
\label{W}
\end{eqnarray}
where  the color indexes are given by the Latin letters and $T\exp (\dots)$ means the 
time ordered exponent. It was proven (see e.g.~\cite{brown1979}) that Eq.~(\ref{W})
can be reduced to the Wilson loop for the colorless $Q\bar Q$ state. 
The corresponding 
Wilson loop is going along the rectangular contour 
which is shown in Fig.~\ref{Wilsonloop}.
At $T\rightarrow\infty$ one may neglect by contributions of 
short sides.

According to Eq.~(\ref{F}), Eq.~(\ref{W}) can be rewritten in 
the operator form as
\begin{eqnarray}
 {W}&=&\int D\xi\exp\left[\frac{1}{2}\sum_{i, j=1}^2\left(\frac{\delta}
{\delta a_a^{(i)}}S_{ab}^{(ij)}(\xi)\frac{\delta}{\delta a_b^{(j)}}\right)\right] 
\cr
&\times&\left. \frac{1}{D^{(1)}-ga^{(1)}}
\frac{1}{\bar D^{(2)}-g\bar a^{(2)}}\right|_{a=0},
\label{W1}
\end{eqnarray}  
where the operator  $D^{(1)}$ is defined as $D^{(1)}=\theta^{-1}-gA^{(1)}(\xi)$ and, 
$a^{(1)}$ and $A^{(1)}$ are the corresponding fields projections to the line $L_{1}$,
respectively. Similarly, one has
 $\bar D^{(2)}=\theta^{-1}-g\bar A^{(2)}(\xi)$ where  
$\bar a^{(2)}$ and  $\bar A^{(2)}$ are the corresponding fields  
projections to the line $L_{2}$,
respectively. 
The lowest order ${\cal O}(\alpha_s\lambda^{1/2})$
matrix element of $W$ is given by Eq.~(\ref{1gluonexchange}), corresponding to the first diagram of 
Eq.~(\ref{diagramsQQ}).

Formal expression corresponding to Eq.~(\ref{W1})
is given by
\begin{eqnarray}
  {W} &=& \int D\xi
\Big[ 
\left(D^{(1)}-\Sigma^{(1)}\right) 
\left(1-\frac{\lambda_a}{2}\left(D^{(1)}-\Sigma^{(1)}\right)^{-1}\right.\cr
&\times &\left.g^2 S_{ab}^{(12)} \left(\bar D^{(2)}-\bar\Sigma^{(2)}\right)^{-1}\frac{\bar\lambda_b}{2} \right) \cr
&\times&
\left(\bar D^{(2)}-\bar\Sigma^{(2)}\right)
\Big]^{-1}
\label{W2}
\end{eqnarray}
where operator $\Sigma$ is defined by its matrix element Eq.~(\ref{defg2}).
Consequently, Pobylitsa's equation  in the approximation 
${\cal O}(\lambda,\alpha_s\lambda^{1/2})$ has form
\begin{eqnarray}
  {W}^{-1}&=&
  {w}^{(1)-1}   {\bar w}^{(2)-1}-\sum_i \int d\xi_i 
\theta^{(1) -1}\cr
&\times&\!\!\!\!
\left(\frac{1}{ D_i^{(1)}}-\theta^{(1)}\right)\theta^{(1) -1}
\theta^{(2) -1}\left(\frac{1}{\bar D_i^{(2)}}-\theta^{(2)}\right)\theta^{(2) -1}\cr
&-&g^2 \frac{\lambda_a}{2}\frac{\bar\lambda_b}{2} \int D\xi\,
 S^{(12)}_{ab},
\label{W-1}
\end{eqnarray}
where $ {w}^{(1)-1}$ is given by Eq.(\ref{w-1}). 
Note, that 
the similar to Eq.(\ref{w-1})  the equation for $ {\bar w}^{(2)-1}$ also can be 
written. 

In order to get $Q\bar Q$ potential $V(r)$ we have to find 
the asymptotic form of the   $Q\bar Q$  correlator~(\ref{W}) 
at large time $T\to\infty$, given by the expression 
$\exp(-VT)$. Here the corresponding potential $V=V_{\rm dir}+V_{\rm pert}$ contains the direct instanton induced part $V_{\rm dir}$ originated from the second term of Eq.~(\ref{W-1}) and the perturbative 
one-gluon exchange part $V_{\rm pert}$ originated from 
the third term, respectively.
A calculation method of $V_{\rm dir}$ from 
the  Pobylitsa's equation was described 
in Ref.~\cite{Diakonov:1989un} and its explicit form 
is given in the next subsection~\ref{subsec:dir}. 
Similar calculations will lead 
to $V_{\rm pert}$ the final form of which is given in the 
subsection~\ref{subsec:1gluon}.

\subsection{Direct instanton induced singlet potential in ILM}
\label{subsec:dir}

Let us first start from the direct instanton induced 
potential $V_{\rm dir}(r)$. It can be evaluated 
by repeating the calculations presented in Ref.~\cite{Diakonov:1989un}. 
Further analysis performed in
Ref.~\cite{Yakhshiev2018}  showed that 
$V_{\rm dir}(r)$ can be written as 
\begin{equation}
V_{\rm dir}(r)=\frac{4\pi\lambda}{N_c\rho}\,
{\cal I}_{\rm dir}\left(\frac{r}{\rho}\right),
\label{VIdir}
\end{equation}
where ${\cal I}_{\rm dir}(x)$ - dimensionless integral 
of the form
\begin{align}
{\cal I}_{\rm dir}(x)=&\int_0^\infty y^2 dy \int_{-1}^1 dt\bigg[1-\cos\left(\frac{\pi y}{\sqrt{y^2+1}}\right)\cr
&\times\cos\left(\pi\sqrt{\frac{y^2+x^2+2xyt}{y^2+x^2+2xyt+1}}\right)\cr
&-\frac{y+xt}{\sqrt{y^2+x^2+2xyt}}\sin\left(\frac{\pi y}{\sqrt{y^2+1}}\right)\cr
&\times\sin\left(\pi\sqrt{\frac{y^2+x^2+2xyt}{y^2+x^2+2xyt+1}}\right)\bigg].
\label{Idirnumer}
\end{align}
At the small distances ($x\ll 1$), ${\cal I}_{\rm dir}(x)$ can be evaluated
analytically and one has the potential
\begin{eqnarray}
V_{\rm dir}(r)&\simeq &\frac{4\pi\lambda}{N_c\rho}\,\left\{\frac{\pi^2}{3}\left[\frac{\pi}{16}-J_1(2\pi)\right]\frac{r^2}{\rho^2}\right.\cr
&-&\!\left.\pi\left[\frac{\pi^2(438+7\pi^2)}{30720}+\frac{J_2(2\pi)}{80}\right]\frac{r^4}{\rho^4}+\dots\!
\right\}\!,\qquad
\end{eqnarray}  
in terms of the Bessel functions $J_{n}$. At the large values of the 
$Q\bar{Q}$ inter-distance ($r\gg \rho$),
the potential has the form 
\begin{equation}
V_{\rm dir}(r)\simeq 2\Delta M_Q^{\rm dir}-
\frac{2\pi^3\lambda}{N_c r}.
\label{VNPlargex}
\end{equation}

\begin{figure}[hbt]
\vskip 0.5cm
\includegraphics[width=7cm,angle=0]{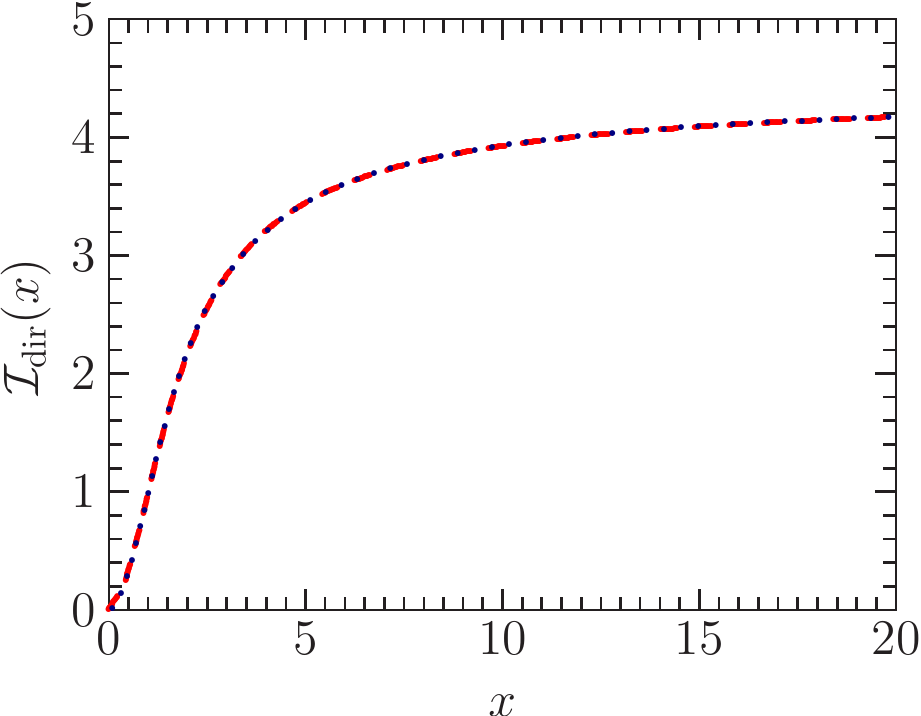}
\caption{(Color online) The dimensionless integral ${\cal I}_{\rm dir}(x)$ from the instanton vacuum. The numerical result of Eq.\,(\ref{Idirnumer}) is depicted as the red dashed curve whereas that of parametrization given in
 Eq.\,(\ref{Idirpar}) is drawn as the blue dotted curve. }
\label{VCdir}
\end{figure}

The function ${\cal I}_{\rm dir}(x)$  can be calculated  numerically or 
parametrized with the very high precision
as it was done in Ref.~\cite{Yakhshiev:2018juj}.
However, it also can be fitted in the  
Gaussian form as
\begin{eqnarray}
V_{\rm dir}(r)&=&\frac{4\pi\rho^3}{N_cR^4}\,
{\cal I}_{\rm dir}\left(\frac{r}{\rho}\right)\,,\\
{\cal I}_{\rm dir}(x)&=&{\cal I}_0^{\rm d}\left\{1+
\sum_{i=1}^2\left[a_i^{\rm d}x^{2(i-1)}+a_3^{\rm d}
(-b_3^{\rm d}x)^i\right]e^{-b_i^{\rm d}x^2}\right.\cr
&+&\left.\frac{a_3^{\rm d}}{x}\left(1-
e^{-b_3^{\rm d}x^2}\right)\right\}.
\label{Idirpar}
\end{eqnarray}
depending on the practical purposes.
The parameters corresponding to this parametrization 
have the values
\begin{align}
{\cal I}^{\rm d}_0&=4.41625,\cr 
a_1^{\rm d}&=-1,\quad
a_2^{\rm d}=0.128702,\quad a_3^{\rm d}=-1.1047,\\
b_1^{\rm d}&=0.404875,\quad
b_2^{\rm d}=0.453923,\quad b_3^{\rm d}=0.420733.\nonumber
\end{align}
The comparison of numerical and parametrized forms of
${\cal I}_{\rm dir}(x)$ are shown in Figure~\ref{VCdir}.
One can see almost the one to one correspondence of the numerical 
calculations and the  parametrization Eq.\,(\ref{Idirpar}).

\subsection{Perturbative one-gluon exchange singlet potential in ILM}
\label{subsec:1gluon}

%We may assume that in ILM one-gluon exchange perturbative %$Q\bar Q$ potential
%is given by
%\bea
%V_{\rm pert}(\vec r) 
%=g^2 \frac{\lambda_a}{2}\frac{\bar\lambda_a}{2}\int\frac{d^3k}{(2\pi)^3}\exp(i\vec k\vec r)\frac{1}{\vec k^2+M_g^2(\vec k)}
%\label{Vpert2}
%\eea

\begin{figure}[h]
%\vspace{0.5cm}
\includegraphics[width=7cm,angle=0]{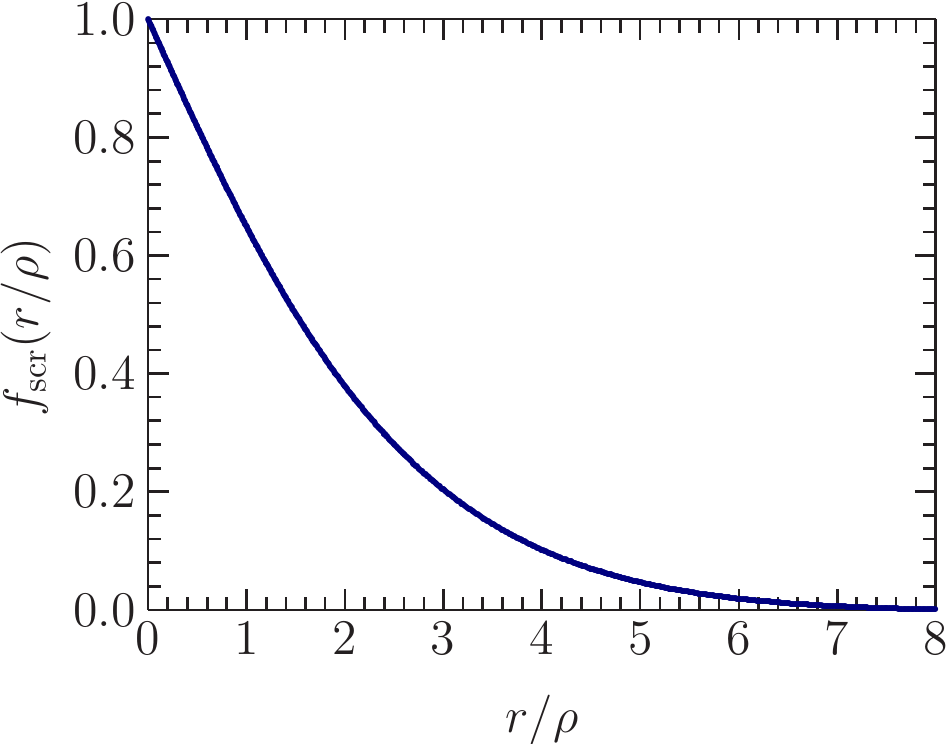}
\caption{(Color online) Dependence of screening function 
$f_{\rm scr}$ on $r/\rho$ at ILM parameters
$ R=1~{\rm fm}$ and $\rho=1/3~{\rm fm}$.}
\label{fscreen}
\end{figure}

From Eq.~(\ref{Vpert1}) it is seen that 
the perturbative one-gluon exchange potential 
has the form
\begin{equation}
V_{\rm pert}=g^2\left(\frac{\lambda_a}{2}\frac{\bar\lambda_b}{2}\right)
\int d\tau\bar S_{ab}(\vec r,\tau).
\end{equation}
Using  the momentum representation of the averaged gluon propagator 
in Eq.\,(\ref{glprop}) and the singlet color factor~(\ref{colorfactor}) from Appendix~\ref{appendix}, one can get
\begin{equation}
V_{\rm pert}(r)=-\frac{4}{3}\,
g^2\int\frac{d^3q}{(2\pi)^3}\frac{e^{i\vec q\cdot\vec r}}{q^2+M_g^2(q)}\,.
%=Cg^2\int\frac{d^3k }{( 2\pi)^3\rho}
%\frac{e^{i\vec k\cdot\vec r/\rho}}{k^2+(\rho M_g(0)k K_1(k))^2}
\label{1gl}
\end{equation}
where $r$ is a distance between $Q$ and $\bar Q$ quarks and $M_g(q)$ is given by Eq.~(\ref{Mg0}).
After the angular integration and introducing the 
dimensionless variables, $x=r/\rho$ and $y=q\rho$, one can obtain 
the perturbative one-gluon exchange potential  in the form of
\begin{align}
\label{Vpertup}
V_{\rm pert}(r)&=-\frac{4\alpha_s}{3r}f_{\rm scr}(x),\\
f_{\rm scr}(x)&=1-\frac{2x}{\pi}
\int_0^\infty 
{\rm d}y~j_0(xy)
\frac{3\pi^2
\lambda{ K}_1^2(y)}{1+3\pi^2
\lambda{ K}_1^2(y)},
\label{Vpert}
\end{align}
where $f_{\rm scr}(x)$ plays the role of screening function (see Fig.~\ref{fscreen}).

\begin{figure*}[hbt]
%\vspace{0.5cm}
\includegraphics[width=7cm,angle=0]{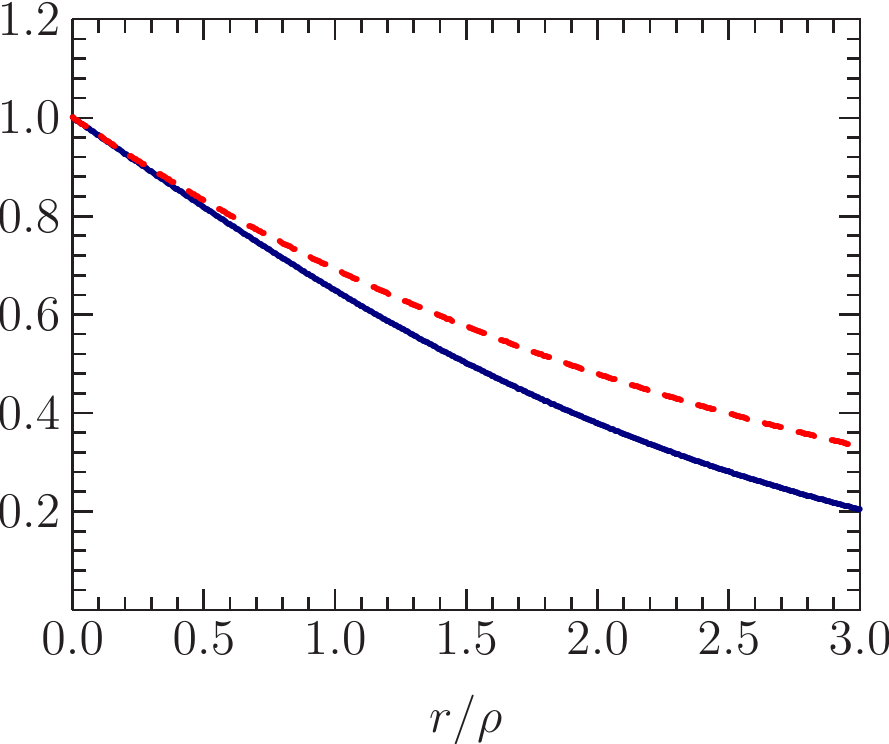}\hskip 1.5cm
\includegraphics[width=7cm,angle=0]{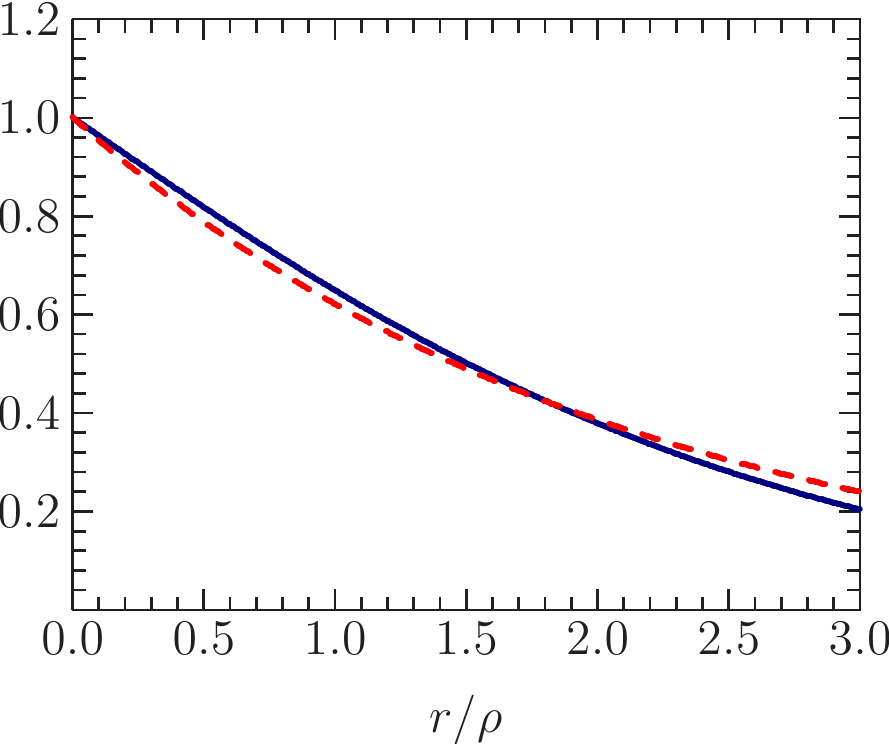}
\caption{(Color online) Dependence of $\exp(-rM_Y)$ (left panel)
and $\exp(-rM_{Y,1})$ (right panel) on $r/\rho$ (orange 
color) in comparison with the exact numerical result 
for $V_{\rm pert}(r/\rho)$ (blue color)
at ILM parameters $ R=1~{\rm fm}$ and $\rho=1/3~{\rm fm}$.
The effective gluon mass must be different at different regions 
of $r$ (see the text) and their  values in the left and right panels 
are chosen as 
$M_Y=218$\,MeV  and $M_{Y,1}=282$\,MeV,
respectively.}
\label{Yukawa}
\end{figure*}

Naturally, from the the Eq.~(\ref{Vpert}) at the short distances one can obtain the Coulomb-like potential 
\begin{align}
V_{\rm pert}(r)&\simeq -\frac{4\alpha_s}{3}
\left(\frac{1}{r}- A\right)\,,\cr
\qquad A &=
\frac{2}{\pi\rho}\,\int_0^\infty dy\,\frac{3\pi^2
\lambda{ K}_1^2(y)}{1+3\pi^2
\lambda{ K}_1^2(y)}\,.
\label{VQQpersmallx}
\end{align}
It is easy to understand the meaning of $A$ comparing with Yukawa-like potential 
$V_Y$ corresponding to the constant gluon mass $M_Y$. 
Its general expression and the behavior at $r\to 0$ are given as 
\begin{align}
V_{Y}(r)&=-\frac{4}{3}\,
g^2\int\frac{d^3q}{(2\pi)^3}\frac{e^{i\vec q\cdot\vec r}}{q^2+M_Y^2}\cr
&=  -\frac{4\alpha_s}{3r}\exp(-M_Y r)\cr
&\to -\frac{4\alpha_s}{3}\left(\frac{1}{r}-M_Y\right)+
...
\label{yukawa}
\end{align}
In such a way,  $M_Y=A$ and
$M_Y$ must be smaller than  
$M_g(0)\simeq 358\,{\rm MeV}$ 
since the influence of form-factor in $M_g(q)$.
From Eq.~(\ref{VQQpersmallx}) it is seen that 
 $M_Y\simeq 218$~ {\rm MeV} at $\rho=1/3$~fm, $R=1$~fm. 
One can see from the left panel of Fig.~\ref{Yukawa},
that at small $r<0.5 \rho$ there is coincidence of 
$V_Y$ in Eq.\,(\ref{yukawa}) with the exact numerical 
result (see Eq.\,(\ref{Vpert})) as it was expected. 
But at the large distances $r>0.5\rho$ 
the slope of curve corresponding to the 
exact solution is different and 
$V_Y$ must be 
described by another parameter, i.e. $M_{Y,1}\approx 282$~MeV 
(see the right panel in Fig.~\ref{Yukawa}). 
This conclusion is natural since the large distance $r$ corresponds 
to the smaller momentum $q$ and one should have 
$M_Y<M_{Y,1}<M_g(0)$.
The parametrization of $V_{\rm pert}$ in the form of Yukawa 
potential may be useful for the quick and crude applications.

For the high accuracy calculations one can parametrize
the potential $V_{\rm pert}(r)$ in a much better way.
For that purpose we introduce the dimensionless integral
${\cal I}_{\rm pert}(x)$ in the expression of screening 
function $f_{\rm scr}$ (see Eq.\,(\ref{Vpert}))
in such a way
\begin{align}
f_{\rm scr}(x)=1-\frac{2x}{\pi}\,
{\cal I}_{\rm scr}(x).
\end{align}
Then ${\cal I}_{\rm scr}$ can be parametrized 
with the high precision analogously to ${\cal I}_{\rm dir}$ as 
\begin{align}
{\cal I}_{\rm scr}(x)&={\cal I}_0^{\rm s}\left\{
\sum_{i=1}^2\left[a_i^{\rm s}x^{2(i-1)}+a_3^{\rm s}
(-b_3^{\rm s}x)^i\right]e^{-b_i^{\rm s}x^2}\right.\cr
&+\left.\frac{a_3^{\rm s}}{x}\left(1-
e^{-b_3^{\rm s}x^2}\right)\right\}.
\label{Iscrpar}
\end{align}
The parameters corresponding to this parametrization 
have the values
\begin{align}
{\cal I}^{\rm s}_0&=0.578695,\cr
a_1^{\rm s}&=1,\quad
a_2^{\rm s}=0.121348,\quad a_3^{\rm s}=2.71619,\cr
b_1^{\rm s}&=0.144123,\quad
b_2^{\rm s}=0.189758,\quad b_3^{\rm s}=b_1^{\rm s}
\end{align}
for the ILM parameters $\rho=1/3$\,fm and $R=1$\, fm.\footnote{Note,
that ${\cal I_{\rm dir}}$ is ILM parameters independent parametrization while
the parametrization ${\cal I}_{\rm scr}$ depends on the ratio $\rho/R$.} 
For another set of parameters,  $\rho=0.36$\,fm and $R=0.89$\,fm, one 
has the following parameters
\begin{align}
{\cal I}^{\rm s}_0&=0.743964,\cr
a_1^{\rm s}&=1,\quad
a_2^{\rm s}=-0.0362548,\quad a_3^{\rm s}=2.11812,\cr
b_1^{\rm s}&=0.329341,\quad
b_2^{\rm s}=0.460567,\quad b_3^{\rm s}=b_1^{\rm s}.
\end{align}

\begin{figure}[hbt]
\vskip 0.5cm
\includegraphics[width=7cm,angle=0]{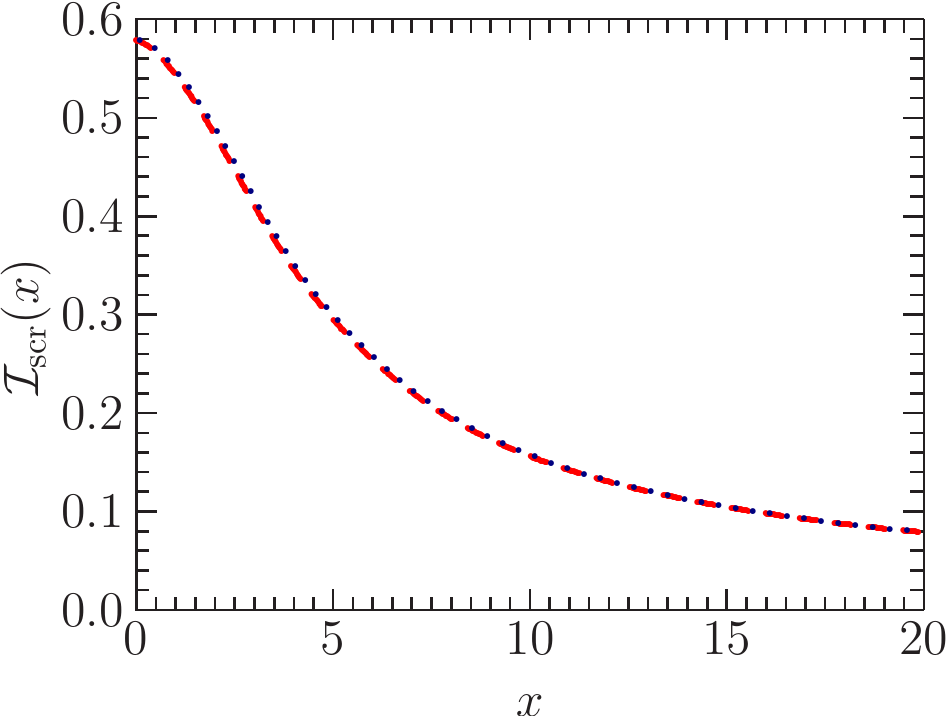}
\caption{(Color online) The dimensionless integral ${\cal I}_{\rm scr}(x)$ from the instanton vacuum
including the perturbative corrections. The numerical result is depicted as the red dashed curve whereas that of parametrization Eq.\,(\ref{Iscrpar}) is drawn as the blue dotted curve. The ILM parameters are set as $\rho=1/3$\,fm and $R=1$\,fm.}
\label{Iscr}
\end{figure}

One can see that the relations ${\cal I}_0^{\rm s}a_3^{\rm s}\simeq\pi/2$ is hold. 
%\com{(I think it must be 
%$a_3^{\rm s}=e$ if we evaluate analytically. I didn't try yet.)}
The comparison of numerical and parametrized forms of 
${\cal I}_{\rm scr}(x)$ is shown in Figure~\ref{Iscr}.
Like in the case of ${\cal I}_{\rm dir}$, here also we have almost the one-to-one correspondence.

At large distances $r>\rho$ 
the potential $V_{\rm pert}(r)$  is not long-ranged anymore 
and quickly goes to zero. In such a way, at large distances 
the instanton medium produces the screening effect 
in the one gluon exchange perturbative potential.

\section{Order of instanton effects}
\label{sec:InsEffects}

Obviously, the best and straightforward  way of 
estimation of the instanton effects is 
the analysis of charmonia and bottmonia states and 
comparison the results in ILM approach with the results of other 
phenomenological approaches.   For that purpose one
needs the spin-dependent parts of the $Q\bar{Q}$ potential.
Although the relations between the central and spin-dependent parts of the direct instanton contributions
are already known, one should re-calculate such
relations in the case of one gluon exchange 
perturbative interaction which includes the instanton 
contributions. Such a detailed analysis we will leave for
the future work and concentrate here only on the central
part of $Q\bar{Q}$ interactions.

In the present work, we will ignore the spin splitting 
effects in the charmonium spectra and concentrate only 
on some low lying S-wave states. First, we perform the fully 
variational calculations considering the different 
sets of potentials without and with the instanton effects.
The next, we consider 
the instanton effects in the first order  pertubation theory
and compare with the results of the fully variational
calculations. The corresponding conclusions 
from these studies will be helpful in our future works. 

Further, we define the full
central $Q\bar{Q}$ potential which includes all possible instanton effects in the following form 
\begin{equation}
V(r)=\sigma r + V_{\rm pert}(r)+V_{\rm dir}(r),
\end{equation}
which supplies the confinement phenomena at large distances.
This potential leads to the standard Cornell's potential
Eq.\,(\ref{cornell}) in the absence of instanton effects.
In order to account the instanton effects in the first 
order perturbation theory we divide the Hamiltonian
into two parts 
\begin{equation}
H=H_0+\tilde H,
% =-\frac{1}{m_Q}\vec\nabla^2+V,
\label{FullH}
\end{equation} 
where $H_0$ is Hamiltonian of the Cornell's model and 
$\tilde H$ is the perturbative part of Hamiltonian. 
They are defined as
\begin{align}
H_0&=-\frac{1}{m_Q}\vec\nabla^2+V_{\rm Cornell},\cr
\tilde H&= V-V_{\rm Cornell} \equiv V_{\rm dir}+V_{\rm scr},\cr
 V_{\rm scr}&=-\frac{4\alpha_s}{3}
(f_{\rm scr}-1).
\end{align}

The results of full variational 
calculations are presented in 
Table~\ref{Table1} for the possible two sets 
of ILM parameters. The details of calculations can 
be found in Ref.\,\cite{Yakhshiev:2018juj}. As an example 
of the Cornell's model parameters, we chosed the 
parameter set MWOI presented in Table\,I of
Ref.\,\cite{Yakhshiev:2018juj}. 
For comparison, in Table\,\ref{Table1}
we present the results for 
Cornell's potential,``Cornell + instanton'' potentials which have 
the nature of instanton contributions from the different regions
and also for the full potential which takes into account all 
possible instanton effects from the different regions. 
\setcounter{table}{0}
\begin{table*}[hbt]
%\label{Table1}
\caption{The results of full variational calculations.
 It is considered 
only some of the  S-wave states corresponding to 
the charmonium states (Spin dependent parts of 
interactions are not included).  The general 
parameters are set as $m_Q=1275$\,MeV and $\sigma=0.17$\,GeV$^2$.
% and  $\alpha_s=0.2$.
Other ILM and the  potential related parameters are given in the table.
The first column is radial 
excitations and the other columns are the results 
corresponding to the different 
sets of potentials. All states 
are given in units of MeV.
}
\centering
\begin{tabular}{|c|c|c|c|c|c|c|c|}
\hline
&&\multicolumn{3}{c|}{$\rho=1/3$\,fm, $R=1$\,fm}&
\multicolumn{3}{c|}{$\rho=0.36$\,fm, $R=0.89$\,fm}\\
\cline{3-8}
$n$& $V_{\rm Cornell}$ & 
$V_{\rm Cornell}+V_{\rm scr}$& $V_{\rm Cornell}+V_{\rm dir}$& $V$
& $V_{\rm Cornell}+V_{\rm scr}$& $V_{\rm Cornell}+V_{\rm dir}$& $V$\\
\hline
\multicolumn{8}{|c|}{$\alpha_s=0.2$}\\
\hline
 1& 3069 &3129 &3111 &3172 &3131 &3146 &3208\\
 2& 3611 &3664 &3682 &3736 &3661 &3747 &3798\\
 3 & 4035 &4079 &4119 &4163&4078 &4198 &4241\\
 4 & 4405 &4443&4496&4534&4442 &4582 &4621\\
 \hline
\multicolumn{8}{|c|}{$\alpha_s=0.4$}\\
\hline
 1& 2905&3026&2943&3063&3031&2973&3100\\
 2& 3509&3617&3579&3688&3612&3642&3747\\
 3& 3956&4044&4038&4128&4043&4116&4205\\
 4& 4337&4414&4428&4505&4413&4514&4591\\
 \hline
\end{tabular}
\label{Table1}
\end{table*}

We note that the both potentials $V_{\rm scr}$ and 
$V_{\rm dir}$ are positively defined and, therefore, 
give the positive contributions to  the whole spectrum. 
One can see the corresponding results 
in the Table\,\ref{Table1}.
Our results are presented for the two values of 
strong coupling constant, 
$\alpha_s=0.2$ and $\alpha_s=0.4$ in order 
to investigate $\alpha_s$ dependence. Obviously, the
increasing value of $\alpha_s$ leads to the strengthening 
of Coulomb-like attraction and lowers the states.

The order of instanton contributions is not big but 
they are not negligible too.  One may conclude, that the 
corresponding instanton effects may be 
considered as the perturbative corrections. 
In order to 
understand this situation better we will consider the 
first order perturbative corrections to the Cornell's model 
results considering the instanton effects as the small 
perturbations. 

The comparisons of the corresponding 
perturbative and variational 
calculations are shown in Table~\ref{Table2}.
\begin{table*}[hbt]
\caption{The perturbative vs full variational calculations.
The first column is radial exitations, 2-4 columns are 
the first order perturbative corrections, 5-7 columns are 
the corresponding differences of variational calculations with and without instanton 
generated potentials,
respectively (see explanations in the text).
The parameters and other definitions are same 
as in the Table\,\ref{Table1}.}
\centering
\begin{tabular}{|c|r|r|c|r|r|c|}
\hline
&\multicolumn{3}{c|}{First order perturbative corrections}&
\multicolumn{3}{c|}{The corresponding variational calculations}\\
\cline{2-7}
$n$ & $V_{\rm scr}$&  $V_{\rm dir}$& $V_{\rm scr}+V_{\rm dir}$ &``$V_{\rm scr}$''&  ``$V_{\rm dir}$''& ``$V_{\rm scr}+V_{\rm dir}$''\\
\hline
\multicolumn{7}{|c|}{$\alpha_s=0.2$, $\rho=1/3$\,fm, $R=1$\,fm, $\lambda=0.01235$}\\
\hline
1&60.124&44.305&104.430&60.119&42.439&102.611\\
2&52.826&72.224&125.050&52.707&71.438&124.651\\
3&43.864&84.342&128.206&43.743&83.873&127.954\\
4&38.247&91.518&129.765&38.172&91.193&129.561\\
\hline
\multicolumn{7}{|c|}{$\alpha_s=0.2$, $\rho=0.36$\,fm, $R=0.89$\,fm, $\lambda=0.02677$}\\
\hline
1&61.661&82.116&143.777&61.607&77.142&139.449\\
2&50.557&138.826&189.383&50.451&136.343&187.599\\
3&43.144&163.750&206.893&43.047&168.332&205.956\\
4&37.891&178.749&216.640&37.820&177.813&216.013\\ \hline
\multicolumn{7}{|c|}{$\alpha_s=0.4$, $\rho=1/3$\,fm, $R=1$\,fm, $\lambda=0.01235$}\\
\hline
1&120.343&39.088&159.432&120.311&37.369&157.725\\
2&108.071&70.832&178.902&107.662&69.962&178.584\\
3&89.237&83.728&172.965&88.733&83.231&172.671\\
4&77.342&91.216&168.558&77.030&90.877&168.317\\
\hline
\multicolumn{7}{|c|}{$\alpha_s=0.4$, $\rho=0.36$\,fm, $R=0.89$\,fm, , $\lambda=0.02677$}\\
\hline
1&125.739&72.026&197.765&125.593&67.620&194.298\\
2&102.992&135.925&238.918&102.579&133.184&237.412\\
3&87.474&162.424&249.898&87.085&160.914&249.191\\
4&76.523&178.078&254.601&76.226&177.090&254.114\\
\hline
\end{tabular}
\label{Table2}
\end{table*}
In the left-half of the table, we present 
the first order perturbative corrections due to instantons
(see $\tilde H$ in Eq.\,(\ref{FullH}))
calculated on a basis of Cornell's
model wave functions corresponding to the Hamiltonian $H_0$.  
On the right-half of the table, we present
the corresponding differences of variational calculations with and without instanton 
generated potentials. For example, 
``$V_{\rm scr}$" means the difference between the results 
of the potential models, ``$V_{\rm Cornell}+V_{\rm scr}$'' 
and ``$V_{\rm Cornell}$", obtained by means the variational 
calculations (The corresponding results  are 
presented in Table\,\ref{Table1}.). 
It should be compared with the 
first order perturbative corrections corresponding to the 
perturbation potential $V_{\rm scr}$. 
One can see that almost the hundred percent effects due 
to instantons can be considered as the first order perturbative
corrections to the spectrum.

When the value of $\alpha_s$ is changed, the general 
picture will not change if one concentrates to the order of 
instanton contributions, i.e. they still remain as the 
first order perturbative corrections. 
It can be seen from the upper-half and the lower-half parts in 
Table\,\ref{Table2}.
The relative sizes of all possible instanton effects 
in comparison with the results corresponding to the 
Cornell's model results found to
be from 3\,\% to 6\,\% depending on the parameters of 
instanton 
liquid model and the excitation state, e.g. compare the column 
$V_{\rm Cornell}$ in Table\,\ref{Table1} with the column 
$V_{\rm scr}+V_{\rm dir}$ in Table\,\ref{Table2}. 

From other side, it is interesting to see that
the contribution amount itself increases 
two times if we concentrate on $V_{\rm scr}$ value
(compare the upper-half and  the lower-half parts of Table\,
\ref{Table2}). This is obvious result while the parameter
$\alpha_s$ is overall factor in the expression 
of $V_{\rm scr}$ in Eq.\,(\ref{Vpertup}).
In contrast, $V_{\rm dir}$ is independent on
parameter $\alpha_s$ and depends on it only by the
wave function changes which are small 
(again compare the upper-half and 
the lower-half parts of Table\,
\ref{Table2}).

It is also interesting to note, that the increasing value 
of packing parameter $\lambda$ will not change much
the contribution  from screening part $V_{\rm scr}$.
This can be seen by comparisons of $V_{\rm scr}$ or 
``$V_{\rm scr}$'' values for the fixed value of $\alpha_s$
(see the upper-half or the lower-half part of Table\,
\ref{Table2}).  This is due to the fact that 
$f_{\rm scr}$ depends on $\lambda$ 
in a nontrivial way (see Eq.\,(\ref{Vpert})). In contrast, 
the contribution from the direct instanton  interactions 
increases almost linearly which is 
seen from the columns  $V_{\rm dir}$ or 
``$V_{\rm dir}$'' in Table\,\ref{Table2}. One can 
note, that an approximately
two times increased value of $\lambda$ increases the 
value of direct instanton interaction contributions also 
approximately 
two times (again see the upper-half or 
the lower-half part of 
Table\,\ref{Table2}). 
This can be seen also from the expression 
of $V_{\rm dir}$ where $\lambda$ is overall 
factor (see Eq.\,(\ref{VIdir})).\footnote{Note, 
also that the another overall factor $1/\rho$ is almost
 same for  the both values of $\rho$, $1/3$\,fm and 
$0.36$\,fm. }  

Although the direct comparisons of results with the 
experimental data can be done 
after inclusions of the spin-dependent parts in the 
potential, here we already can make some qualitative 
predictions. For that purpose, first we will refer to the 
Tables\, I and \,II presented in Ref.\,\cite{Yakhshiev:2018juj}.
One can see, that with the value of $\alpha_s=0.2068$
denoted as MWOI it is possible 
to fit the experimental data by using 
Cornell's type potential and concentrating on the 
first six low-lying S-wave states during the fitting process
(see Table\,II in Ref.\,\cite{Yakhshiev:2018juj} and 
explanations there).
While the low-lying states are reproduced more or less
well the excited states are lower estimated. Inclusion
of the direct instanton interactions improves much 
better the low-lying states but the excited states 
are overestimated (see columns M-I and M-IIb in
Table\,II of  Ref.\,\cite{Yakhshiev:2018juj}). 

Now, let us concentrate back to the Table\,\ref{Table2} 
in present work. From the column $V_{\rm dir}$
or ``$V_{\rm dir}$'' on can see that the direct instanton 
interactions will contribute more and more to the excited 
stated. It is remarkable, that the inclusion of screening 
effect from the instantons changes the situation 
drastically. This is due to the fact that the screening 
effect softens the contributions to excited states from 
the instantons (see columns $V_{\rm scr}$
or ``$V_{\rm scr}$''). As a result, the instanton effects 
from both $V_{\rm scr}$ and $V_{\rm dir}$ 
accumulated in such a way that the ground state
is changed in a different way while the excited 
states have more or less overall shift effect
(see columns $V_{\rm scr}+V_{\rm dir}$
or ``$V_{\rm scr}+V_{\rm dir}$''). This situation may 
be quite helpful in describing the experimental data
related to the charmonium states by using the potential 
approach in the framework of instanton liquid model. 

Summarizing all our discussions we note, that  
the instanton effects are at the level of few percents.
Nevertheless, one cannot ignore the instanton effects in the 
heavy quarkonium spectra. They may also be   
important during the fine tuning processes of the whole 
spectrum 
which takes into account the spin-spin, spin-orbit and tensor 
interactions. 

\section{Summary and Outlook}
\label{sec:SumOut}

In this work we studied one body and two body correllators 
corresponding to the 
heavy quark sector in the framework of Instanton Liquid Model
by inclusion also the perturbative corrections. 
Although the instantons cannot explain 
a confinement mechanism,  we
have shown that they play a nontrivial role not only 
in the nonperturbative region $r\sim\rho$ but also in the 
perturbative region $r<\rho$. In the perturbative 
region the instantons will affect in such way that the one gluon 
exchange perturbative potential becomes the short range 
and remains the screened at large distances.
In such a way, at short distances the OGE potential including the 
instanton effects can be approximated by a Yukawa-type potential 
with the corresponding dynamical gluon mass playing the role of 
parameter in the exponent. 
 The direct 
instanton effects reproduce an overall shift at the 
nonperturbative region which can be accounted as the 
renormalization of heavy quark mass in the instanton medium.
Consequently, our conclusion is that the instanton 
effects in both, perturbative and nonperturbative, 
regions are important for understanding the heavy 
quark physics.

At quantative level, we also estimated the instanton 
effects to the whole 
spectrum and found out that they can completely 
be considered as the first order perturbative corrections. 
The relative size of instanton effects in the spectra of heavy 
quarkonia at the level of few percents and should be taken 
into account during the fine tuning processes.
In Ref.\,\cite{Yakhshiev:2018juj} it was discussed that 
the direct instanton effects from the nonperturbative region 
may explain the origin of phenomenological potential 
parameters fitted to the spectrum of heavy quarkonia.  
The inclusion of instanton effects in the perturbative 
region may help not only in the qualitative understanding 
$Q\bar{Q}$ interactions but also at the 
quantitave level concerned on more accurate description 
of spectra. For that  purpose, one should take into 
account the changes in the spin-dependent potentials   
due to instanton effects in the perturbative region.
This can be done by means of relating the central and 
spin-dependent potential in the framework of heavy quark 
effective theory. The corresponding studies are currently 
under the way.

\section*{Acknowledgments}

The work is supported by Uz grant OT-F2-10
(M.M. and N.R.) and by the Basic Science Research Program 
through the National Research Foundation (NRF) of Korea funded 
by the Korean government (Ministry of Education, Science and
Technology, MEST), Grant Number~2020R1F1A1067876 (U.T.Y.).
 
\appendix
\section{Detailed calculation of correlators}
\label{appendix}

In this appendix we discuss the details of perturbative expansions 
for the quark and $Q\bar Q$ correlators, and the corresponding ILM 
contributions to the heavy quark mass and $Q\bar Q$ potential.

\subsection{Contribution to the heavy quark mass in ILM }
\label{subsec:pertcontHQM}

We begin with the expansion of the propagator in 
Eq.\,(\ref{w}). First we note, that the non-averaged in the instanton 
medium heavy quark propagator  in operator notations\footnote{Note, 
that one can also define an 
analogous operator in a single instanton field in the form of 
$D_i=\theta^{-1}-g A_i(\xi_i)$.} $D=\theta^{-1}-g A(\xi)$ 
is given by the expansion
\begin{eqnarray}
&&\left.\exp\left[\frac{1}{2}
\left(\frac{\delta}{\delta a}S\frac{\delta}{\delta a}\right)\right] 
\frac{1}{ D-ga }  \right|_{a=0}\cr
&&~=
\sum_{m=0}^{\infty}\frac{1}{m!}\left[\frac{1}{2}\left(\frac{\delta}{\delta a}S\frac{\delta}{\delta a}\right)\right]^m \left(\frac{1}{D}ga\right)^{2m}\frac{1}{D}
\nonumber\\
&&~=\sum_{m=0}^{\infty}\left(g^2\frac{1}{D}
\left(S\frac{1}{D}\right)\right)^m\frac{1}{D}.
\label{A:Qprop}
\end{eqnarray}
It is easy to find that the lowest order ${\cal{O}}(\alpha_s)$ peturbative correction to the heavy quark  propagator is 
\begin{eqnarray}
&&\frac{1}{D}\left(\frac{\delta}{\delta a}S\frac{\delta}{\delta a}\right)\left(ag\frac{1}{D}\right)^2
\cr
&&~\rightarrow g^2\langle x_0|\frac{1}{D}|x_1\rangle   (S_{ab}(x_1,x_2)+S_{ba}(x_2,x_1))\cr
&&~\times\frac{\lambda_a}{2}\langle x_1|\frac{1}{D}|x_2\rangle \frac{\lambda_b}{2}\langle x_2|\frac{1}{D}|x_3\rangle
\cr
&&~\equiv g^2\langle x_0|\frac{1}{D}|x_1\rangle   \langle x_1|(S\frac{1}{D})|x_2\rangle \langle x_2|\frac{1}{D}|x_3\rangle  ,
\label{g2}\end{eqnarray}
where repeating $x_i$ means integration over $x_i$.
Also,
 $g^2(S\frac{1}{D})$ has a meaning of the perturbative heavy quark mass operator $\Sigma$ in the lowest order $O(\alpha_s)$ and was defined by its matrix element
\begin{eqnarray}
\langle x_1|\Sigma|x_2\rangle &=&
g^2\langle x_1|(S\frac{1}{D})|x_2\rangle \cr
&\equiv& g^2(S_{ab}(x_1,x_2)+S_{ba}(x_2,x_1))\cr
&\times&\frac{\lambda_a}{2}
\langle x_1|\frac{1}{D}|x_2\rangle \frac{\lambda_b}{2}.
\label{defg2}
\end{eqnarray}
Analogously, the next order contribution ${\cal{O}}(\alpha_s^2)$ 
has form
\begin{eqnarray}
&&\frac{1}{D}\left(\frac{\delta}{\delta a}S\frac{\delta}{\delta a}\right)^2\left(ag\frac{1}{D}\right)^4\cr
&&~\rightarrow g^4\langle x_0|\frac{1}{D}|x_1\rangle  
\left(\frac{\delta}{\delta a(y)}S(y,z)\frac{\delta}{\delta a(z)} \right)\cr
&&~\times\left(\frac{\delta}{\delta a(y')}S(y',z')\frac{\delta}{\delta a(z')}\right)
\cr
&&~\times\left(a(x_1)\langle x_1|\frac{1}{D}|x_2\rangle a(x_2)\langle x_2|\frac{1}{D}|x_3\rangle \right)\cr
&&~\times\left(a(x_3)\langle x_3|\frac{1}{D}|x_4\rangle a(x_4)\langle x_4|\frac{1}{D}|x_5\rangle \right)\cr
&&~\times\langle x_0|\frac{1}{D}|x_1\rangle \langle x_1|S(x_1,x_2)+S(x_2,x_1)|x_2\rangle \cr
&&~\times \langle x_2|\frac{1}{D}|x_3\rangle \langle x_3|S(x_3,x_4)+S(x_4,x_3)|x_4\rangle \cr
&&~\times\langle x_4|\frac{1}{D}|x_5\rangle +[(x_{i_1},x_{i_2})\to (x_{j_1},x_{j_2})].\nonumber
\end{eqnarray}
We define the sum of all irreducible diagrams as $\Sigma$, which has the following form in diagram representation
\begin{eqnarray}
\label{massoperator}
\raisebox{1ex}{$\Sigma$ =}~
% g^2
\tikz{\begin{feynman}\diagram*{
 i1 -- i2,
%{[edges={fermion}]i0 -- i1 -- i2 -- i3},
i1 -- [gluon, half left] i2;
};
\end{feynman}}
\quad\raisebox{1ex}{+}\quad
%g^4
\tikz{\begin{feynman}\diagram*{
 i1 -- i2 -- i3 -- i4 ,
%{[edges={fermion}]i0 -- i1 -- i2 -- i3 -- i4 -- i5},
i1 -- [gluon, half left] i3,
i2 -- [gluon, half left] i4;
};
\end{feynman}}
\quad\raisebox{1ex}{+}\raisebox{1ex}{...}
\end{eqnarray}
Now, the series~(\ref{A:Qprop}) can be summed up 
according to the geometrical progression
\begin{eqnarray}
\frac{1}{D}\left(1+\Sigma\frac{1}{D}+\Sigma\frac{1}{D}\Sigma\frac{1}{D}+\cdots\right)&=&\frac{1}{D}\frac{1}{1-\Sigma\frac{1}{D}}\cr
&=&\frac{1}{D-\Sigma}.
\end{eqnarray}
However, as we already mentioned in the text, we neglected by 
the gluon self-interaction terms  ${\cal{O}}(g^3,g^4)$. 
Therefore, we have to take into account the heavy quark mass operator 
only in the lowest order ${\cal{O}}(\alpha_s)$ single-loop 
approximation. 
That is $\Sigma = g^2(S D^{-1})$ defined by Eq.~(\ref{defg2}). 

Consequently, we obtain the following expression for the non-averaged heavy quark propagator
\begin{eqnarray}
\frac{1}{ D -g^2(S D^{-1})}.
\label{HQprop}
\end{eqnarray}
Further we averaging Eq.~(\ref{HQprop}) over the instanton ensemble 
in ${\cal{O}}(\lambda,\alpha_s\lambda^{1/2})$ approximation, 
which leads to
further simplification of the perturbative mass operator as 
$g^2(S D^{-1})\approx g^2(S\theta)$. 

It is obvious, that in order to have the ILM perturbative heavy quark 
mass operator we have to 
remove from $\Sigma$ the perturbative heavy quark mass operator in 
the empty space as
$\Sigma-\Sigma_0$, where $\Sigma_0 = g^2(S^0 \theta)+O(g^4).$  
Then, 
Eq.~(\ref{HQprop}), should be rewritten as
\begin{eqnarray}
\left[ \theta^{-1} -g\sum_i A_i(\xi_i) -g^2\left(\sum_j \Delta S^j(\xi_j)\theta\right)\right]^{-1}.\quad
\label{Qprop}
\end{eqnarray}
%%%%%%%%% end of propagator expansion %%%%%%%%%%%%%

To calculate ILM direct and ILM perturbative mass contributions we use more definitive form of  Pobylitsa Eq.~(\ref{w-1}) as
\begin{eqnarray}
&&\langle t_2|w^{-1}|t_1\rangle  =-\delta^{'}(t_2-t_1)\cr
&&~+\sum_{\pm}\int dz_{4,\pm} f(t_2-z_{4,\pm},t_1-z_{4,\pm})\cr
&&~+g(t_2,t_1),
\label{mew1}
\end{eqnarray}
where
\begin{eqnarray}
&&f(t_2-z_{4,\pm},t_1-z_{4,\pm})\cr
&&~=-\frac{N}{2V}\sum_{\pm}\tr_c \int d^3 z_\pm \, 
\langle t_2|\theta^{-1}(w_\pm-\theta)\theta^{-1}|t_1\rangle 
\cr
&&g(t_2,t_1)=-g^2 \langle t_2|\left((\bar S-S^0 )\theta\right)|t_1\rangle .
\end{eqnarray}
In the last equations
$w_{\pm}\equiv (\theta^{-1} - A_\pm(\xi_\pm))^{-1}$ 
and `$\pm$' indexes correspond to the  instanton/antiinstanton, 
respectively. Introducing the Fourier transformation~\cite{Diakonov:1989un}
\begin{eqnarray}
\langle t_2|w|t_1\rangle& =&
\int \frac{ d\omega}{2\pi}\exp[i\omega(t_2-t_1)]w(\omega)\cr
&=&\int \frac{ d\omega}{2\pi}\exp[i\omega(t_2-t_1)]\frac{1}{w^{-1}(\omega)}
\label{fourier1}
\end{eqnarray}
 and using  Pobylitsa Eq.~(\ref{mew1}) one can find that
\begin{eqnarray}
\langle t_2|w|t_1\rangle& =&
\int \frac{ d\omega}{2\pi}\exp[i\omega(t_2-t_1)]\frac{1}{i\omega+f(\omega)+g(\omega)}\nonumber\cr
f_\pm(\omega)&=&\int  dt_2\, dt_1
\exp[-i\omega t_2+i\omega t_1]\cr
&\times&f(t_2-z_{4,\pm},t_1-z_{4,\pm})_{|_{z_{4,\pm}=0}}\\
g(\omega)&=&\int d\tau \exp[i\omega\tau] g(\tau),~\tau=t_2-t_1. 
\label{fourier2}
\end{eqnarray}
One can see, that there is a pole within the approximations 
discussed above
$$
i\omega_0+f(0)+g(0)=0.
$$

According to this pole we obtain ILM direct and ILM perturbative contributions to the heavy quark mass from the time dependence of the heavy quark propagator Eq.~(\ref{fourier2}) as
\begin{eqnarray}
\langle t_2|w|t_1\rangle& \sim&\exp(-\Delta M_Q(t_2-t_1)),\cr
\Delta M_Q&=&\Delta M_Q^{\rm dir}+\Delta M_Q^{\rm pert},\,\,\,
\nonumber\\
\Delta M_Q^{\rm dir}&=&f(0),\cr
\Delta M_Q^{\rm pert}&=&g(0).
\label{fourier3}
\end{eqnarray}
Note, that the direct instanton contributions
to the heavy quark mass
$\Delta M_Q^{\rm dir}$ was calculated in Ref.~\cite{Diakonov:1989un}. 

Taking into account the color factor in 
$((\bar S-S^0)\theta)$, which is  $\frac{\lambda_a}{2}\frac{\lambda_b}{2}\delta_{ab}=\frac{4}{N_c}I$ ($I$ is $N_c\times N_c$ unit matrix)
we have ILM perturbative heavy quark mass contribution
in the following form
 \begin{eqnarray}
\Delta M_Q^{\rm pert}I&=& - g^2\int_{-\infty}^{\infty} d(t_2-t_1) 
\langle t_2|\left((\bar S-S^0) \theta\right)|t_1\rangle\cr 
&=&-g^2\frac{ \lambda_a}{2}\frac{ \lambda_b}{2}
\int_{0}^{\infty} dt (\bar S_{ab}(t)-S_{ab}^0(t))\cr
&=&\frac{2}{N_c}I\,g^2\int \frac{d^3k}{(2\pi)^3}
\frac{M_g^2(k)}{k^2\left[k^2+M_g^2(k)\right]}\cr
&\le& \frac{2}{N_c}I\alpha_s M_g(0).\nonumber
\end{eqnarray}
One can estimate, that
\begin{eqnarray}
\Delta M_Q^{\rm pert}&\le& \frac{2}{N_c}\alpha_s M_g(0)\sim  M_Q^{\rm dir}\sim 70\, {\rm MeV} 
\label{DMpert1}
\end{eqnarray} 
at the values of parameters
  $N_c=3$, $\alpha_s= 0.3,\,\rho=1/3\, {\rm fm}$, $R=1\, {\rm fm}.$  
 %%%%%%%%%%%%%%%%%%%%%%%%%%%%%%%%%%%%%%%%%%%%%%%%%%
\subsection{Perturbative contribution to the heavy quark - antiquark potential in ILM }
\label{subsec:pertpot}

We will proceed with the $Q\bar Q$ correlator 
Eq.~(\ref{W1}) in an analogous way what we did with the
heavy quark propagator.
 
%================ expansion of Q bar Q correlator ======================
Consequently, for the non-averaged $Q\bar Q$ correlator Eq.~(\ref{W1}) one can write the expansion
\begin{eqnarray}\label{A:QbarQ}
&&\exp\left[\frac{1}{2}\sum_{i\neq j=1}^2\left(\frac{\delta}
{\delta a^{(i)}}S^{(ij)}(\xi)\frac{\delta}{\delta a^{(j)}}\right)\right] \cr
&&~\times\left.\frac{1}{D^{(1)}-ga^{(1)}}
\frac{1}{\bar D^{(2)}-g\bar a^{(2)}}\right|_{a=0}\cr
&&~=\sum_{m=0}^\infty\frac{1}{m!}\left(\frac{g^2}{2}\right)^m\left[\sum_{i\neq j=1}^2\left(\frac{\delta}
{\delta a^{(i)}}S^{(ij)}(\xi)\frac{\delta}{\delta a^{(j)}}\right)\right]^m\cr
&&~\times\left(\frac{1}{D^{(1)}}a^{(1)}\right)^m\frac{1}{D^{(1)}}\left(\frac{1}{\bar D^{(2)}}\bar a^{(2)}\right)^m
\frac{1}{\bar D^{(2)}}.
\end{eqnarray} 
For the lowest order  ${\cal{O}}(\alpha_s)$ term we obtain
the explicit expression
\begin{eqnarray}
&&\frac{g^2}{2}\int dx_2^{(1)}dx_2^{(2)}\cr
&&\times\langle x^{(1)}_3|\frac{1}{D^{(1)}}|x_2^{(1)}\rangle \frac{\lambda_a}{2}S_{ab}^{(12)}(x^{(1)}_2,x^{(2)}_2)\langle x_2^{(1)}|\frac{1}{D^{(1)}}|x^{(1)}_1\rangle \cr
&&\times\langle x^{(2)}_1|\frac{1}{\bar D^{(2)}}|x_2^{(2)}\rangle \frac{\bar\lambda_b}{2}\langle x_2^{(2)}|\frac{1}{\bar D^{(2)}}|x^{(2)}_3\rangle \cr
&&+\langle x^{(1)}_3|\frac{1}{D^{(1)}}|x_2^{(1)}\rangle \frac{\lambda_b}{2}S_{ba}^{(21)}(x^{(2)}_2,x^{(1)}_2)\langle x_2^{(1)}|\frac{1}{D^{(1)}}|x^{(1)}_1\rangle \cr
&&\times\langle x^{(2)}_1|\frac{1}{\bar D^{(2)}}|x_2^{(2)}\rangle \frac{\bar\lambda_a}{2}\langle x_2^{(2)}|\frac{1}{\bar D^{(2)}}|x^{(2)}_3\rangle. 
\end{eqnarray}
Accordingly, the lowest ${\cal O}(\alpha_s\lambda^{1/2})$ contribution to $W$ is given by
\begin{eqnarray}
&&g^2\int dx_2^{(1)}dx_2^{(2)}
 \langle x^{(1)}_3|\theta^{(1)}|x_2^{(1)}\rangle \langle x_2^{(1)}|\theta^{(1)}|x^{(1)}_1\rangle 
 \cr
&&~\times \frac{\lambda_a}{2}\frac{\bar\lambda_b}{2}
 \bar S_{ab}(x^{(1)}_2 - x^{(2)}_2)\cr
&&~\times \langle x^{(2)}_1|\theta^{(2)}|x_2^{(2)}\rangle \langle x_2^{(2)}|\theta^{(2)}|x^{(2)}_3\rangle ,
 \label{lhs2}
\end{eqnarray}
which correspond to the first diagram in the following 
series of irreducible diagrams~(\ref{diagramsQQ})
\begin{eqnarray}
&& %first part
\tikz{\begin{feynman}
\vertex (j0);
\vertex [right=1cm of j0] (j1);
\vertex [right=1cm of j1] (j2);
\vertex [below=1cm of j0](i0);
\vertex [right=1cm of i0] (i1);
\vertex [right=1cm of i1] (i2);
\diagram*{
j0 -- j1 -- j2,
i0 -- i1 -- i2,
%{[edges={anti fermion}]j0 -- j1 -- j2},
%{[edges={fermion}]i0 -- i1 -- i2},
i1 -- [gluon] j1;
};
\vertex[below=1em of i0] {\(x_1^{(1)}\)};
\vertex[below=1em of i1] {\(x_2^{(1)}\)};
\vertex[below=1em of i2] {\(x_3^{(1)}\)};
\vertex[above=1em of j0] {\(x_1^{(2)}\)};
\vertex[above=1em of j1] {\(x_2^{(2)}\)};
\vertex[above=1em of j2] {\(x_3^{(2)}\)};
\end{feynman}}
~\raisebox{1.1cm}{+}~
\tikz{\begin{feynman}
\vertex (j0);
\vertex [right=1cm of j0] (j1);
\vertex [right=1cm of j1] (j2);
\vertex [right=1cm of j2] (j3);
\vertex [below=1cm of j0](i0);
\vertex [right=1cm of i0] (i1);
\vertex [right=1cm of i1] (i2);
\vertex [right=1cm of i2] (i3);
\diagram*{
j0 -- j1 -- j2 -- j3,
i0 -- i1 -- i2 -- i3,
%{[edges={anti fermion}]j0 -- j1 -- j2},
%{[edges={fermion}]i0 -- i1 -- i2},
i1 -- [gluon] j2,
i2 -- [gluon] j1,
};
\vertex[below=1em of i0] {\(x_0^{(1)}\)};
\vertex[below=1em of i1] {\(x_1^{(1)}\)};
\vertex[below=1em of i2] {\(x_2^{(1)}\)};
\vertex[below=1em of i3] {\(x_3^{(1)}\)};
\vertex[above=1em of j0] {\(x_0^{(2)}\)};
\vertex[above=1em of j1] {\(x_1^{(2)}\)};
\vertex[above=1em of j2] {\(x_2^{(2)}\)};
\vertex[above=1em of j3] {\(x_3^{(2)}\)};
\end{feynman}}
~\raisebox{1.1cm}{+}~~
\raisebox{1.1cm}{...}
\label{diagramsQQ}
\end{eqnarray}
 
In the approximation ${\cal{O}}(\alpha_s\lambda^{1/2})$, 
one should sum up according 
to the geometrical progression and make 
a ladder by repetition of only the first diagram in~(\ref{diagramsQQ}).
We expect that this diagram will provide the perturbative contribution to 
the  $Q\bar Q$ potential in the approximation ${\cal{O}}(\alpha_s\lambda^{1/2})$. The corresponding matrix element is given by
\begin{eqnarray}
&&g^2\frac{\lambda_a}{2}\frac{\bar\lambda_b}{2}
\int_{t_1}^{t_3}\int_{t_1}^{t_3}  dt_2^{(1)}dt_2^{(2)}
\bar S_{ab}(\vec r, t^{(1)}_2-t^{(2)}_2)
\cr
&&~=(t_3-t_1)g^2 \frac{\lambda_a}{2}\frac{\bar\lambda_a}{2}\int\!\!\frac{d^3k}{(2\pi)^3}\frac{e^{i\vec k\vec r}}{\vec k^2+M_g^2(\vec k)}.\qquad
\label{1gluonexchange}
\end{eqnarray}
In the operator form one has
\begin{eqnarray}
W&=& 
\int D\xi\frac{1}{ D^{(1)}}\frac{1}{ D^{(2)}}
+g^2 \theta^{(1)} \theta^{(2)}\cr
&&\times
\frac{\lambda_a}{2}\frac{\bar\lambda_b}{2}
 \bar S_{ab}
\theta^{(1)} \theta^{(2)}\,+...
\label{operator1gluonexchange}\\
\bar S_{ab}&=&\int D\xi\,S_{ab}^{(12)}.
\nonumber
\end{eqnarray}
Here  the matrix element of operator $S^{(12)}$ is given by
\begin{align}
&\langle x^{(1)}_2,x^{(2)}_2|S^{(12)}|x^{(1)}_1,x^{(2)}_1\rangle\cr
&~=
S^{(12)}(x^{(2)}_1,x^{(1)}_1)\delta^4(x^{(2)}_2-x^{(2)}_1)\cr
&~\times\delta^4(x^{(1)}_2-x^{(1)}_1).
\end{align}
One has the similar expression for $S^{(21)}$, while 
\begin{eqnarray}
&&\langle x^{(1)}_2,x^{(2)}_2|\bar S|x^{(1)}_1,x^{(2)}_1\rangle\cr
&&~ =
\bar S(x^{(2)}_1-x^{(1)}_1)\delta^4(x^{(2)}_2-x^{(2)}_1)\cr
&&~\times\delta^4(x^{(1)}_2-x^{(1)}_1).
\end{eqnarray}
Operators $S^{(12)}$ and $S^{(21)}$ do not commute with $D^{(1)}$ and 
$ D^{(2)}$ operators
since they are acting in both subspaces. 
Formal expression for the operator $W$ is given by Eq.~(\ref{W2})
and  Pobylitsa equation in the approximation ${\cal O}(\lambda,\alpha_s\lambda^{1/2})$ is 
given by Eq.~(\ref{W-1}), where we will use 
\begin{eqnarray}
\nonumber
\int D\xi\,S_{ab}^{(12)}(x^{(1)}_2,x^{(2)}_2)=\bar S_{ab}(x^{(1)}_2-x^{(2)}_2).
\end{eqnarray}
Asymptotically the $Q\bar Q$ correlator at large time $T\to\infty$ 
is given by $\exp(-VT) $ and
can be calculated in accordance with
the approach used in the Eqs.~(\ref{fourier1})-(\ref{fourier3})~\cite{Diakonov:1989un}. 
Consequently,  the $Q\bar Q$ potential also can be written as
\begin{equation}
V=V_{\rm dir}+V_{\rm pert}.\nonumber
\end{equation}
As it is shown in Ref.~\cite{Diakonov:1989un}
the direct instanton induced $Q\bar Q$ potential 
$V_{\rm dir}$ is originated from the matrix elements of 
the first and second terms in the Eq.~(\ref{W-1}).
It is explicitly expressed as
\bea
&&V_{\rm dir}= -\int dt^{(1)}_2\,dt^{(2)}_2\, dt^{(1)}_1\,dt^{(2)}_1
\sum_i\int d\xi_i \langle x^{(1)}_2\,x^{(2)}_2|\cr
&&~\times
\theta^{(1) -1}\theta^{(2) -1}\left(\frac{1}{ D_i^{(1)}}\frac{1}{ D_i^{(2)}}-\theta^{(1)}\theta^{(2)}\right)\cr
&&~\times\theta^{(1) -1}\theta^{(2) -1}
|x^{(1)}_1\,x^{(2)}_1\rangle.
\label{Vdir}
\eea
For the one-gluon exchange  $Q\bar Q$  potential in ILM
  $V_{\rm pert}$ we have the similar relation
\bea
V_{\rm pert}&=&\int dt^{(1)}_2\,dt^{(2)}_2\, dt^{(1)}_1\,dt^{(2)}_1 \frac{\lambda_a}{2}\frac{\bar\lambda_b}{2} \cr
&&~\times\left(S^0_{ab}+ \sum_i\int d\xi_i \Delta S_{ab}\right)
(\vec r,t^{(1)}_1,t^{(2)}_1)
\cr
&&~\times\delta(t^{(1)}_1-t^{(1)}_2)\delta(t^{(2)}_1-t^{(2)}_2)\cr
&&~=g^2 \frac{\lambda_a}{2}\frac{\bar\lambda_b}{2}\int d\tau \bar S_{ab}(r,\tau),\\
&&\tau=t^{(1)}_1-t^{(2)}_1.\nonumber
\label{Vpert1}\eea
%%%%%%%%% end of Q bar Q expansion %%%%%%%%%%%%%%%
%=========== color factor =======================
For the color factor in Eq.~(\ref{Vpert1}) 
we have the following expressions (see~\cite{brown1979})
\begin{eqnarray}
\frac{\lambda_a}{2}\frac{\lambda_a}{2}=
 \frac{\bar\lambda_a}{2}\frac{\bar\lambda_a}{2}=\frac{N_c^2-1}{2N_c}I,\,\,\,
 \bar\lambda_a=-\lambda_a^T,
\cr
I_a=\frac{\lambda_a}{2}+\frac{\bar\lambda_a}{2},\,\,\,
I_aI_a=\frac{N_c^2-1}{N_c}I +2\frac{\lambda_a}{2}\frac{\bar\lambda_a}{2}.
\eea
One has $(I_aI_a)_S=0$ in the color singlet state  and 
$(I_aI_a)_A=N_c$ in the adjoint state, respectively. Finally, one has
\begin{eqnarray}
\left(\frac{\lambda_a}{2}\frac{\bar\lambda_a}{2}\right)_{S}&=&-\frac{N_c^2-1}{2N_c}I,\cr
\left(\frac{\lambda_a}{2}\frac{\bar\lambda_a}{2}\right)_{A}
&=&\frac{1}{2N_c}I.
\label{colorfactor}
\end{eqnarray}


\begin{thebibliography}{99}

%\cite{Eichten:1974af}
\bibitem{Eichten:1974af} 
  E.~Eichten, K.~Gottfried, T.~Kinoshita, J.~B.~Kogut, K.~D.~Lane and T.~M.~Yan,
``The Spectrum of Charmonium'',
  Phys.\ Rev.\ Lett.\  {\bf 34}, 369 (1975)
  Erratum: [Phys.\ Rev.\ Lett.\  {\bf 36}, 1276 (1976)].
%  doi:10.1103/PhysRevLett.34.369, 10.1103/PhysRevLett.36.1276

\bibitem{Bali:2000gf} 
  G.~S.~Bali,
 ``QCD forces and heavy quark bound states'',
  Phys.\ Rept.\  {\bf 343}, 1 (2001).
%  doi:10.1016/S0370-1573(00)00079-X
%  [hep-ph/0001312].
  %%CITATION = doi:10.1016/S0370-1573(00)00079-X;%%

%\cite{Brambilla:2009bi}
\bibitem{Brambilla:2009bi}
N.~Brambilla, A.~Vairo, X.~Garcia Tormo, i and J.~Soto,
``The QCD static energy at NNNLL'',
Phys. Rev. D \textbf{80} (2009), 034016.
%doi:10.1103/PhysRevD.80.034016
%[arXiv:0906.1390 [hep-ph]].

%\cite{Mateu:2018zym}
\bibitem{Mateu:2018zym}
V.~Mateu, P.~G.~Ortega, D.~R.~Entem and F.~Fernández,
``Calibrating the Naïve Cornell Model with NRQCD'',
Eur. Phys. J. C \textbf{79} (2019) 323.
%doi:10.1140/epjc/s10052-019-6808-2
%[arXiv:1811.01982 [hep-ph]].

%\cite{Diakonov:2002fq}
\bibitem{Diakonov:2002fq} 
  D.~Diakonov,
  ``Instantons at work'',
  Prog.\ Part.\ Nucl.\ Phys.\  {\bf 51} (2003) 173.
%  doi:10.1016/S0146-6410(03)90014-7
% [hep-ph/0212026].
  %%CITATION = doi:10.10

%\cite{Schafer:1996wv}
\bibitem{Schafer:1996wv} 
  T.~Schafer and E.~V.~Shuryak,
  ``Instantons in QCD'',
  Rev.\ Mod.\ Phys.\  {\bf 70}  (1998) 323.
%  doi:10.1103/RevModPhys.70.323
%  [hep-ph/9610451].
 
%\cite{shuryak2018}
\bibitem{shuryak2018} 
  Edward Shuryak,
  ``Lectures on nonperturbative QCD ( Nonperturbative Topological Phenomena in QCD and Related Theories)'',
[arXiv:1812.01509 [hep-ph]].

%\cite{Goeke:2007bj}
\bibitem{Goeke:2007bj}
  K.~Goeke, M.~M.~Musakhanov and M.~Siddikov,
 ``Low energy constants of chi PT from the instanton vacuum model'',
  Phys.\ Rev.\ D {\bf 76} (2007) 076007.
%  doi:10.1103/PhysRevD.76.076007
%  [arXiv:0707.1997 [hep-ph]].\\
  %%CITATION = doi:10.1103/PhysRevD.76.076007;%%

%\cite{Goeke:2007nc}
\bibitem{Goeke:2007nc}
  K.~Goeke, H.~C.~Kim, M.~M.~Musakhanov and M.~Siddikov,
  ``1/N(c) corrections to the magnetic susceptibility of the QCD vacuum'',
  Phys.\ Rev.\ D {\bf 76} (2007) 116007.
  %doi:10.1103/PhysRevD.76.116007
%  [arXiv:0708.3526 [hep-ph]].\\
  %%CITATION = doi:10.1103/PhysRevD.76.116007;%%
  
%\cite{Goeke:2010hm}
\bibitem{Goeke:2010hm}
  K.~Goeke, M.~Musakhanov and M.~Siddikov,
  ``QCD isospin breaking ChPT low-energy constants from the instanton vacuum'',
  Phys.\ Rev.\ D {\bf 81} (2010) 054029.
  %doi:10.1103/PhysRevD.81.054029
%  [arXiv:1002.0283 [hep-ph]].\\
  %%CITATION = doi:10.1103/PhysRevD.81.054029;%%
  
%\cite{Musakhanov:2012zm}
\bibitem{Musakhanov:2012zm}
  M.~Musakhanov,
  ``QCD in Infrared Region and Spontaneous Breaking of the Chiral Symmetry'',
  PoS Baldin{\bf-ISHEPP-XXI} (2012) 008.
  %doi:10.22323/1.173.0008
%  [arXiv:1212.0947 [hep-ph]].\\
  %%CITATION = doi:10.22323/1.173.0008;%%
  
%\cite{Musakhanov:2018sdu}
\bibitem{Musakhanov:2018sdu}
  M.~Musakhanov,
  ``Gluons, Heavy and Light Quarks in the QCD Vacuum'',
  EPJ Web Conf.\  {\bf 182} (2018) 02092.
  %doi:10.1051/epjconf/201818202092
%  [arXiv:1802.06211 [hep-ph]].
  %%CITATION = doi:10.1051/epjconf/201818202092;%%
  
 \bibitem{FJR1976} L.D. Faddeev, ``Looking for multi-dimensional solitons"
in: \textit{Non-local Field Theories}, Dubna, 1976.

%\cite{Jackiw:1976pf}
\bibitem{Jackiw:1976pf}
R.~Jackiw and C.~Rebbi,
``Vacuum Periodicity in a Yang-Mills Quantum Theory'',
Phys. Rev. Lett. \textbf{37} (1976) 172.
%doi:10.1103/PhysRevLett.37.172
 
 %\cite{Belavin:1975fg}
\bibitem{Belavin:1975fg}
A.~Belavin, A.~M.~Polyakov, A.~Schwartz and Y.~Tyupkin,
``Pseudoparticle Solutions of the Yang-Mills Equations'',
Phys. Lett. B \textbf{59} (1975) 85.
%doi:10.1016/0370-2693(75)90163-X
 
 %\cite{Chu:1994vi}
\bibitem{Chu:1994vi}
M.~Chu, J.~Grandy, S.~Huang and J.~W.~Negele,
``Evidence for the role of instantons in hadron structure from lattice QCD'',
Phys. Rev. D \textbf{49} (1994) 6039.
%doi:10.1103/PhysRevD.49.6039
%[arXiv:hep-lat/9312071 [hep-lat]].

%\cite{Negele:1998ev}
\bibitem{Negele:1998ev}
J.~W.~Negele,
``Instantons, the QCD vacuum, and hadronic physics'',
Nucl. Phys. B Proc. Suppl. \textbf{73} (1999) 92.
%doi:10.1016/S0920-5632(99)85010-5
%[arXiv:hep-lat/9810053 [hep-lat]].

%\cite{DeGrand:2001tm}
\bibitem{DeGrand:2001tm}
T.~A.~DeGrand,
``Short distance current correlators: Comparing lattice simulations to the instanton liquid'',
Phys. Rev. D \textbf{64} (2001) 094508.
%doi:10.1103/PhysRevD.64.094508
%[arXiv:hep-lat/0106001 [hep-lat]].

%\cite{Faccioli:2003qz}
\bibitem{Faccioli:2003qz}
P.~Faccioli and T.~A.~DeGrand,
``Evidence for instanton induced dynamics, from lattice QCD,''
Phys. Rev. Lett. \textbf{91} (2003) 182001.
%doi:10.1103/PhysRevLett.91.182001
%[arXiv:hep-ph/0304219 [hep-ph]].

%\cite{Millo:2011zn}
\bibitem{Millo:2011zn}
R.~Millo and P.~Faccioli,
``Computing the Effective Hamiltonian of Low-Energy Vacuum Gauge Fields'',
Phys. Rev. D \textbf{84} (2011) 034504.
%doi:10.1103/PhysRevD.84.034504
%[arXiv:1105.2163 [hep-ph]].

%\cite{Kraan:1998kp}
\bibitem{Kraan:1998kp}
T.~C.~Kraan and P.~van Baal,
``Exact T duality between calorons and Taub - NUT spaces'',
Phys. Lett. B \textbf{428} (1998) 268.
%doi:10.1016/S0370-2693(98)00411-0
%[arXiv:hep-th/9802049 [hep-th]].

%\cite{Kraan:1998pm}
\bibitem{Kraan:1998pm}
T.~C.~Kraan and P.~van Baal,
``Periodic instantons with nontrivial holonomy'',
Nucl. Phys. B \textbf{533} (1998) 627.
%doi:10.1016/S0550-3213(98)00590-2
%[arXiv:hep-th/9805168 [hep-th]].
%366 citations counted in INSPIRE as of 28 May 2020

%\cite{Lee:1998bb}
\bibitem{Lee:1998bb}
K.~M.~Lee and C.~h.~Lu,
``SU(2) calorons and magnetic monopoles'',
Phys. Rev. D \textbf{58} (1998) 025011.
%doi:10.1103/PhysRevD.58.025011
%[arXiv:hep-th/9802108 [hep-th]].
%316 citations counted in INSPIRE as of 28 May 2020

%\cite{Diakonov:2009jq}
\bibitem{Diakonov:2009jq}
D.~Diakonov,
``Topology and confinement'',
Nucl. Phys. B Proc. Suppl. \textbf{195} (2009) 5.
%doi:10.1016/j.nuclphysbps.2009.10.010
%[arXiv:0906.2456 [hep-ph]].


%\cite{Liu:2015ufa}
\bibitem{Liu:2015ufa}
Y.~Liu, E.~Shuryak and I.~Zahed,
``Confining dyon-antidyon Coulomb liquid model. I.'',
Phys. Rev. D \textbf{92} (2015) 085006.
%doi:10.1103/PhysRevD.92.085006
%[arXiv:1503.03058 [hep-ph]].

%\cite{Liu:2015jsa}
\bibitem{Liu:2015jsa}
Y.~Liu, E.~Shuryak and I.~Zahed,
``Light quarks in the screened dyon-antidyon Coulomb liquid model. II.'',
Phys. Rev. D \textbf{92} (2015) 085007.
%doi:10.1103/PhysRevD.92.085007
%[arXiv:1503.09148 [hep-ph]].

%\cite{Digal:2005ht}
\bibitem{Digal:2005ht}
S.~Digal, O.~Kaczmarek, F.~Karsch and H.~Satz,
``Heavy quark interactions in finite temperature QCD'',
Eur. Phys. J. C \textbf{43} (2005) 71.
%doi:10.1140/epjc/s2005-02309-7
%[arXiv:hep-ph/0505193 [hep-ph]].

%\cite{Eichten:1979ms}
\bibitem{Eichten:1979ms}
E.~Eichten, K.~Gottfried, T.~Kinoshita, K.~Lane and T.~M.~Yan,
``Charmonium: Comparison with Experiment'',
Phys. Rev. D \textbf{21} (1980) 203.
%doi:10.1103/PhysRevD.21.203

%\cite{He:1986yq}
\bibitem{He:1986yq}
Y.~He, F.~Wang and C.~W.~Wong,
``Nucleon Core Size and Nuclear Forces'',
Phys. Lett. B \textbf{168} (1986) 177.
%doi:10.1016/0370-2693(86)90959-7

\bibitem{Weise}
 W. Weise, ``Quarks, chiral symmetry and dynamics of nuclear constituents", in:\textit{Quarks and Nuclei},
 %(edited by W. Weise), 
 World Scientific, 1985.
 %, pp. 57-188;
 
 \bibitem{Tegen}
 R.Tegen, ``Nucleon form factors from elastic scattering of polarized
 leptons ($e$, $\mu$, $\tau$) from polarized nucleons" in:\textit{Weak and Electromagnetic Interactions in Nuclei},
%Proceedings... 
%of the International Symposium, Heidelberg, July 1–5, 1986 
%Nucleon form factors...
%(edited by H. V. Klapdor), 
Springer, 1986.
%, pp.435-440.

%\cite{Diakonov:1989un}
\bibitem{Diakonov:1989un}
D.~Diakonov, V.~Petrov and P.~Pobylitsa,
``The Wilson Loop and Heavy Quark Potential in the Instanton Vacuum'',
Phys. Lett. B \textbf{226} (1989) 372.
%doi:10.1016/0370-2693(89)91213-6
 
 %\cite{brown1979} 
 \bibitem{brown1979} 
  L.~S.~Brown and W.~I.~Weisberger,
  ``Remarks on the Static Potential in Quantum Chromodynamics'',
  Phys.\ Rev.\ D {\bf 20} (1979) 3239.


%\cite{Pobylitsa:1989uq}
\bibitem{Pobylitsa:1989uq}
P.~Pobylitsa,
``The Quark Propagator and Correlation Functions in the Instanton Vacuum'',
Phys. Lett. B \textbf{226} (1989) 387.
%doi:10.1016/0370-2693(89)91216-1

 %\cite{Musakhanov:2017erp}
\bibitem{Musakhanov:2017erp}
  M.~Musakhanov and O.~Egamberdiev,
  ``Dynamical gluon mass in the instanton vacuum model'',
  Phys.\ Lett.\ B {\bf 779} (2018) 206.
 % doi:10.1016/j.physletb.2018.01.080
 % [arXiv:1706.06270 [hep-ph]].
  %%CITATION = doi:10.1016/j.physletb.2018.01.080;%%

%\cite{Schafer:1994fd}
\bibitem{Schafer:1994fd}
T.~Schafer and E.~V.~Shuryak,
``Glueballs and instantons'',
Phys. Rev. Lett. \textbf{75} (1995) 1707.
%-1710
%doi:10.1103/PhysRevLett.75.1707
%[arXiv:hep-ph/9410372 [hep-ph]].

%\cite{Tichy:2007fk}
\bibitem{Tichy:2007fk}
M.~C.~Tichy and P.~Faccioli,
``The Scalar glueball in the instanton vacuum'',
Eur. Phys. J. C \textbf{63} (2009) 423.
%doi:10.1140/epjc/s10052-009-1111-2
%[arXiv:0711.3829 [hep-ph]].


%\cite{deForcrand:1991kc}
\bibitem{deForcrand:1991kc}
P.~de Forcrand and K.~F.~Liu,
``Glueball wave functions in lattice gauge calculations'',
Phys. Rev. Lett. \textbf{69} (1992) 245.
%doi:10.1103/PhysRevLett.69.245

%\cite{Weingarten:1994vc,Chen:1994uw}
\bibitem{Weingarten:1994vc}
D.~Weingarten,
``QCD spectroscopy'',
Nucl. Phys. B Proc. Suppl. \textbf{34} (1994) 29.
%-46
%doi:10.1016/0920-5632(94)90318-2
%[arXiv:hep-lat/9401021 [hep-lat]].

%\cite{Chen:1994uw}
\bibitem{Chen:1994uw}
H.~Chen, J.~Sexton, A.~Vaccarino and D.~Weingarten,
``The Scalar and tensor glueballs in the valence approximation'',
Nucl. Phys. B Proc. Suppl. \textbf{34} (1994) 357.
%doi:10.1016/0920-5632(94)90389-1
%[arXiv:hep-lat/9401020 [hep-lat]].

%\cite{Morningstar:1999rf}
\bibitem{Morningstar:1999rf}
C.~J.~Morningstar and M.~J.~Peardon,
``The Glueball spectrum from an anisotropic lattice study'',
Phys. Rev. D \textbf{60} (1999) 034509.
%doi:10.1103/PhysRevD.60.034509
%[arXiv:hep-lat/9901004 [hep-lat]].

%\cite{Athenodorou:2020ani}
\bibitem{Athenodorou:2020ani}
A.~Athenodorou and M.~Teper,
``The glueball spectrum of SU(3) gauge theory in 3+1 dimension'',
[arXiv:2007.06422 [hep-lat]].

%\cite{Meyer:2004jc,Meyer:2004gx}
\bibitem{Meyer:2004jc}
H.~B.~Meyer and M.~J.~Teper,
``Glueball Regge trajectories and the pomeron: A Lattice study'',
Phys. Lett. B \textbf{605} (2005) 344.
%doi:10.1016/j.physletb.2004.11.036
%[arXiv:hep-ph/0409183 [hep-ph]].

%129 citations counted in INSPIRE as of 24 Aug 2020
%\cite{Meyer:2004gx}
\bibitem{Meyer:2004gx}
H.~B.~Meyer,
``Glueball regge trajectories'',
[arXiv:hep-lat/0508002 [hep-lat]].

 
 
%\cite{Yakhshiev2018} 
 \bibitem{Yakhshiev2018} 
 U.~T.~Yakhshiev, H.~C.~Kim, M.~M.~Musakhanov,
E.~Hiyama and B.~Turimov, ``Instanton effects on the heavy-quark static potential'',
 Chin.\ Phys.\ C \textbf{41} (2017) 083102.
 %, {[}arXiv:1602.06074 {[}hep-ph{]}{]}.

\bibitem{Yakhshiev:2018juj} 
  U.~T.~Yakhshiev, H.~C.~Kim and E.~Hiyama,
  ``Instanton effects on charmonium states'',
  Phys.\ Rev.\ D {\bf 98} (2018) 114036.
%  doi:10.1103/PhysRevD.98.114036
%  [arXiv:1811.05608 [hep-ph]].
  %%CITATION = doi:10.1103/PhysRevD.98.114036;%%

 
\end{thebibliography}
\end{document}